\documentclass[epj,nopacs]{svjour}
\usepackage{epsfig,amsfonts,latexsym}

\newcommand{\bbox}{\lower0.85pt\hbox{$\Box$}}
\newcommand{\kreisl}{\raise0.85pt\hbox{$\scriptstyle\bigcirc$}}
\newcommand{\dreieck}{\raise0.85pt\hbox{$\scriptstyle\bigtriangledown$}}
\newcommand{\stern}{\lower0.85pt\hbox{\Large $\star$}}

\journalname{EPJ C}
\begin{document}
\setcounter{page}{1}
\title{
\vspace{-30pt}
{\rm
\rightline{\small UL-NTZ 19/98,
KANAZAWA-98-09}}
\vspace{5pt}
Wave functions and spectrum in hot electroweak matter
for large Higgs masses}
\titlerunning{Wave functions and spectrum in hot electroweak matter}
\authorrunning{E.-M.~Ilgenfritz et al.}
\author{E.-M.~Ilgenfritz \inst{1}\thanks{ilgenfri@hep.s.kanazawa-u.ac.jp} %
\and A.~Schiller \inst{2}\thanks{schiller@tph204.physik.uni-leipzig.de} %
\and C.~Strecha \inst{2}\thanks{strecha@tph204.physik.uni-leipzig.de}}
\institute{Institute for Theoretical Physics, Kanazawa University,
Kanazawa 920-1192, Japan
\and
Institut f\"ur Theoretische Physik, Universit\"at Leipzig,
D-04109 Leipzig, F.R.~Germany}
\date{Received: August 3, 1998}
\abstract{
  We present results for the wave functions and the screening mass spectrum
  for quantum numbers $0^{++}$, $1^{--}$ and $2^{++}$ in the three-dimensional
  $SU(2)$--Higgs model near to the phase transition line below the endpoint
  and in the crossover region. Varying the $3D$ gauge couplings we study the
  behaviour along a line of constant physics towards the continuum limit in
  both phases. In the crossover region the changing spectrum of screening
  states versus temperature is examined showing the aftermath of the phase
  transition at lower Higgs mass. Different to smearing concepts we used large
  sets of operators with various extensions allowing to identify wave
  functions in position space.}

\maketitle

\section{Introduction}

During the last years, due to efforts of three
groups~\cite{kajantie}-\cite{endpoint} using the $3D$ approach, various
aspects of the high temperature electroweak phase transition have been
explored in the $SU(2)$--Higgs model, with increasing Higgs mass $M_H$.  The
interface tension has been seen falling over three orders of magnitude
\cite{kajantie,interface} (if expressed as $\alpha/T_c^3$), whereas the latent
heat goes to zero roughly linearly with $M_H$.  The continuum characteristics
of the transition is difficult to obtain with the same precision by $4D$ Monte
Carlo simulations~\cite{4D}, although the results are consistent with each
other where they can be compared~\cite{rummu,interface}.  Approaching the $W$
mass $M_W$, the character of the transition is now known to change
drastically. It is the spectrum of physical excitations in this parameter
region that we are going to study here.

The $4D$ lattice approach meets two problems which made the $3D$ approach so
attractive.  At first, only the bosonic sector can be studied because of the
problems to deal with chiral fermions.  Second, simulations are difficult due
to the presence of different length scales.  The inverse Matsubara frequency,
$1/(2 \pi T)$, is much shorter than the two correlation lengths, $1/m_W(T)$
and $1/m_H(T)$, with the screening masses differing strongly from the zero
temperature masses $M_W$ and $M_H$.  If the temporal lattice extent $N_{\tau}$
should be not too small, they can be accommodated on a $4D$ lattice only if
this is anisotropic.  This requires more couplings to be tuned than in the
isotropic case when one first has to match the $M_H/M_W$ ratio and the
renormalised coupling $g_4^2(\mu_4)$ on a $T=0$ ($N_{\tau}=N_s$) lattice.

The $3D$ approach represents itself as a radical solution to both problems.
It is based on perturbative dimensional reduction in the continuum combined
with lattice perturbation theory \cite{generic}, leaving to lattice simulation
only the non-perturbative description of the softest modes whose behaviour
governs the transition.  All non-zero Matsubara modes are integrated out
leaving an effective, superrenormalisable theory in $3D$ with the smallest
length scale $1/(2 \pi T)$ removed.  Chiral fermions can be implicitly dealt
with at this step.  In a second step the next heaviest modes related to the
Debye mass $m_D \sim g~T$ (the $A_0$ components of the gauge field) are
integrated out.  Thus an effective $3D$ $SU(2)$--Higgs theory emerges with the
action
\begin{eqnarray}
  S_3 &=& \int d^3 x \Big(\frac{1}{4} F_{\mu \nu}^b F_{\mu
    \nu}^b + (D_{\mu} \phi)^+ (D_{\mu} \phi) 
    \nonumber \\
  & & + m_3^2 \phi^+ \phi + \lambda_3 (\phi^+ \phi)^2 \Big) \, .
  \label{eq:S3Dcontinuum}
\end{eqnarray}
It has dimensionful, renormalisation group invariant couplings $g_3^2$ and
$\lambda_3$ and a running mass squared $m_3^3(\mu_3)$.  It is put into
correspondence to a lattice model with the action
\begin{eqnarray}
  S &=& \beta_G \sum_p \big(1 - \frac{1}{2} \mbox{tr} U_p \big) -  
  \beta_H \sum_{x,\mu}
  \frac{1}{2} \mbox{tr} (\Phi_x^+ U_{x, \mu} \Phi_{x + \hat \mu})
  \nonumber \\
  & & + \sum_x  \big( \rho_x^2 + \beta_R (\rho_x^2-1)^2 \big) \,.
  \label{eq:S3Dlattice}
\end{eqnarray}
The lattice couplings are (with a suitable parameter $M_H^*$)
\begin{eqnarray}
  \beta_G &=& \frac{4}{a g_3^2}\, \, , \, 
  \beta_R = \frac{\lambda_3}{g_3^2} 
  \frac{\beta_H^2}{\beta_G}
  = \frac18 \left( \frac{M_H^*}{80\; \mbox{GeV}}\right)^2 
  \frac{\beta_H^2}{\beta_G}\,, \nonumber  \\ 
  \beta_H&=&\frac{2 (1-2\beta_R)}{6+a^2 m_3^2} \, ,
  \label{eq:couplings}
\end{eqnarray}
which can be expressed in terms of $4D$ couplings and masses.  The parameter
$M_H^*$ is approximately equal to the zero temperature physical Higgs mass.
The summation in (\ref{eq:S3Dlattice}) is taken over plaquettes $p$, sites $x$
and links $l=\{x,\mu\}$.  The gauge fields are represented by unitary $2
\times 2$ link matrices $U_{x,\mu}$ and $U_p$ denotes the $SU(2)$ plaquette
matrix. The Higgs field is parametrised as follows: $\Phi_x = \rho_x V_x$,
where $\rho_x^2= \frac12 \mbox{\rm tr}(\Phi_x^+\Phi_x)$ is the Higgs modulus
squared, and $V_x$ is an element of the group $SU(2)$.

The bare mass squared is related to the renormalised $m_3^2(\mu_3)$ (we choose
$\mu_3=g_3^2$) through a lattice two-loop calculation~\cite{laine} giving
$m_3^2(\mu_3)=m_3^2+ m^2_{\mathrm {1-loop}}+m^2_{\mathrm{2-loop}}$.  The
lattice results obtained in the $3D$ approach indicate the validity of the
dimensional reduction (without additional operators in the action) near to the
transition temperature for Higgs masses in the range between $30$ and $240$
GeV.  The relations between the parameters of the effective $3D$ model and the
physical quantities Higgs mass and temperature are derived in \cite{generic}.

According to recent lattice studies~\cite{endpointh,endpoint,LatticeCrossover}
the standard electroweak theory ceases to possess a first order transition for
a Higgs mass $M_H > 72$ GeV.  Therefore, taking the newest lower bound of the
Higgs mass into account \cite{HiggsMassLimit}, the standard model does not
pass through a true phase transition at the electroweak vector boson mass
scale.  This and the small amount of CP violation in the standard model seem
to rule out the possibility to explain the BAU generation without new physics.
Therefore the phenomenological interest has moved to extensions of the
standard model, with minimal supersymmetric extensions (MSSM) being the most
promising variant.  First interesting lattice results have been published
recently \cite{mssm}.

From the point of view of non-perturbative physics in general, the lattice
version of the standard Higgs model is still interesting as a laboratory for
investigating the behaviour of hot gauge fields coupled to scalar matter, for
the characterisation of possible bound states, for the understanding of real
time topological transitions and the role of embedded topological defects
\cite{CGIS} at the transition. It has been important as a cross-check for
analytical approximation schemes and will be so in future.

We are concentrating in this paper on the qualitative change of the spectrum
of screening states that happens across the phase transition and what remains
of it in the crossover region (at somewhat higher Higgs mass).  This would not
be possible without a systematic evaluation of the wave functions in
configuration space that, for the first time in this context, is attempted
here.  Some intermediate results have been published before \cite{future}.

The rest of this paper is organised as follows.  In Sect.~\ref{sec2} we
discuss the cross correlation technique and the operator set as we have used
them.  The wave functions and masses of the ground state and excited states
near to the endpoint of the phase transition are presented in Sects. 3 and 4
for the symmetric and the Higgs phase at $M_H^*=70$ GeV, respectively. This
complements a recent study in \cite{philipsen} which has been performed,
however, at much lighter ($M_H^*= 35$ GeV) Higgs mass.  Section 5 contains our
results for the dynamics of the spectrum with increasing temperature in the
crossover region for $M_H^*=100$ GeV, slightly above the endpoint of the first
order transition, which can be compared with results of Ref. \cite{philipsen1}
obtained at a markedly larger Higgs mass $M_H^*\approx 120$ GeV.  In Sect. 6
we summarise our results.  The Appendices contain tables of the measured mass
values obtained from this analysis and details how to construct the operator
set for the quantum numbers in $2+1$ dimensions on the lattice.
 

\section{The Cross Correlation Technique in the {\boldmath $3D$} Model}
\label{sec2}

To study simultaneously the ground state {\sl and} excited states (as well as
their wave functions) one has to consider cross correlations between
(time-slice sums of) operators ${\cal {O}}_{i}$ from a complete set in a given
$J^{PC}$ channel with quantum numbers $J$ (angular momentum) , $P$ (parity)
and $C$ (charge conjugation). According to the transfer matrix formalism, one
should be able to write the connected correlation matrix at time\footnote{in
  the $3D$ approach ''time'' is $x_3$} separation $t$ in the spectral
decomposition form
\begin{equation}
  C_{ij}(t) = \sum_{n=1}^{\infty} \Psi_i^{(n)} \Psi_j^{(n)*} e^{-m_n t} 
  \label{eq:spec_dec}
\end{equation}
with
\begin{equation}
  \Psi_i^{(n)}=\langle \mathrm{vac}|
  {\cal{O}}_i | {\bf \Psi}^{(n)}\rangle
\end{equation}
where $|{\bf \Psi}^{(n)}\rangle$ is the $n$-th (zero momentum) energy
eigenstate. The vacuum state is dropped from this sum due to the connectedness
of the correlator.  By suitable diagonalisation this allows to find masses
{\it and} wave functions of the lowest mass screening states (ground state)
and higher mass excited states in the various $J^{PC}$ channels.
 
However, in practice one has to choose a truncated set of operators
${\cal{O}}_i$, ($i=1,\ldots,N$). The hope is that the lowest lying states
($k=1,\ldots,r$, with $r<<N$) might not be essentially affected by the
truncation and that their masses and wave functions can be approximately
extracted from the direct eigenvalue problem for $C_{ij}(t)$
\begin{equation}
  \label{eq:simple_eigen}
  \sum_j  C_{ij}(t) \Psi_j^{(n)} = \lambda^{(n)}(t) \Psi_i^{(n)}. 
\end{equation}
Experience shows that this solution suffers from big systematic truncation
errors.  Solving instead the generalised eigenvalue problem
\begin{equation}
  \sum_j  C_{ij}(t)~\Psi_j^{(n)}  =  
  \lambda^{(n)}(t,t_0) \sum_j C_{ij}(t_0)~\Psi_j^{(n)}   
  \label{eq:gen_eigen_first}
\end{equation}
or
\begin{equation}
  \sum_j  \tilde{C}_{ij}(t,t_0) \tilde{\Psi}_j^{(n)}  =  
  {\lambda}^{(n)}(t,t_0) 
  \tilde{\Psi}_i^{(n)}  \, ,
  \label{eq:gen_eigen}
\end{equation}
with
\begin{equation}
  \tilde{C}_{ij}(t,t_0)= \left( C^{-\frac12}(t_0) C(t) C^{-\frac12}(t_0)
  \right)_{ij}
  \,, 
\end{equation}
($t>t_0$, where practically $t_0=0,1,2,..$) errors related to this truncation
can be kept minimal \cite{luescher,gattringer}.  Practically the decomposition
of the matrix $C(t_0)$ is performed using a Cholesky decomposition $C(t_0)=L
L^T$. The remaining problem is that of diagonalizing a symmetric matrix
$\tilde{C}(t,t_0)=L^{-1} C(t) L^{T~-1}$ with rotated eigenvectors
$\tilde{\Psi}^{(n)}= L^T \Psi^{(n)}$.

The optimised eigenfunctions $\Psi^{(n)}$ in the chosen operator basis
(obtained with a small distance $t_0$) give an information about the overlap
of the source operators ${\cal {O}}_i$ with the actual eigenstates $|{\bf
  \Psi}^{(n)}\rangle$.  Due to truncation the eigenvectors of
(\ref{eq:gen_eigen_first}) are not supposed to be orthogonal to each other
since $C^{-1}(t_0)C(t)$ is not a symmetric matrix. The components of the
eigenvectors $\Psi_k^{(n)}$ ($n \leq N$) are effected by terms of the order
$O({\mathrm{exp}}(-m_{N+1}t))$, where $N$ is the number of used operators.
Only in the limit of a complete set of operators the eigenvectors become
orthogonal.  Therefore, the (non)orthogonality between different states
provides a criterion for the completeness of the operator set.

The masses $m^{(n)}$ of these states are obtained by fitting not the
eigenvalues ${\lambda}^{(n)}$ of (\ref{eq:gen_eigen}) but the diagonal
elements $\mu^{(n)}$,
\begin{equation} 
  \mu^{(n)}(t,t_0)= 
  \sum_{ij}~\tilde{\Psi}_i^{(n)}
  ~\tilde{C}_{ij}(t,t_0)
  ~\tilde{\Psi}_j^{(n)} \, , 
  \label{eq:fit}
\end{equation}
to a hyperbolic cosine form with $t$ in some plateau region of a local
effective mass that is defined as
\begin{equation} 
  m_{\mathrm{eff}}^{(n)}(t,t_0)=\mathrm{log}\frac{\mu^{(n)}(t,t_0)} 
  {\mu^{(n)}(t+1,t_0)} \, .
\end{equation}

The wave function components with respect to the operator basis characterise
the coupling of source operators ${\cal {O}}_i$ to the bound states (lowest
mass or excited) in the $J^{PC}$ channel.  Using operators of different
extension transverse to the correlation direction a spatial resolution of the
optimised wave function can be achieved.  Practically only a subset of
operators with fixed quantum numbers can be included.  The admitted operator
set includes gauge invariant operators properly chosen with respect to the
lattice symmetry and quantum numbers.  For completeness, the choice of the
gauge invariant operators in $2+1$ dimensions on the cubic lattice
corresponding to the quantum numbers mentioned is discussed in Appendix
\ref{quantumnumbers}.  Though the results might be in principle known a
reasonable presentation for the lattice $SU(2)$--Higgs model similar to that
of Ref. \cite{evertz} for 3+1 dimensions was not available.
 
In order to associate a spatial structure to the states under study one has to
use operators which correspond to various extension.  One way is to define a
set of operators using smearing techniques in several variants.  In the
present context this has been practised in Refs.
\cite{teper,philipsen,philipsen1}, smearing gauge links and Higgs fields. The
smearing parameter determines a well-defined mixture of various lengths in the
source operator and has to be optimised by a variational procedure (strictly
speaking, for all states separately).

In contrast to this smearing technique, we have chosen the other extreme and
collected only a few types of operators ${\mathcal{O}}_i$ in our base
(properly chosen with respect to lattice symmetry and quantum numbers) but
with a wide span of sizes $l$ in lattice spacings.  Such a basis allows to
obtain information on the spatial extension of a bound state without going
through a variational procedure (for the lowest and the excited states).  In
principle, smearing the link and Higgs variables underlying these operators
would be conceivable for further optimising the states under study.  Here, we
have restricted ourselves to Higgs strings and Wilson loops with varying
spatial extent.

In the three quantum number channels investigated here we have admitted the
following operators (with $\mu=3$ reserved for the correlation direction, see
Appendix~\ref{quantumnumbers}):
\begin{eqnarray}
  0^{++} :& &  \rho_x^2\nonumber \\
  & &  S_{x,1}(l)+S_{x,2}(l)  \nonumber \\
  & &  W_{x,1,2}(l)+W_{x,2,1}(l) \\
  & &  ({\mathrm{quadratic \ Wilson \ loops \ of \ 
      size}} \  l \times l ) \nonumber \\
  1^{--} :& &  V_{x,1}^b(l)+V_{x,2}^b(l) \nonumber \\
  2^{++} :& &  S_{x,1}(l)-S_{x,2}(l) \nonumber
  \label{extended}
\end{eqnarray}
Here the following notation is used ($l$ denotes the string length in lattice
units)
\begin{eqnarray}
  S_{x,\mu}(l) & = & \frac{1}{2}{\mathrm{tr}}(\Phi^+_x U_{x,\mu}\ldots
  U_{x+(l-1)\hat\mu,\mu}\Phi_{x+l \hat \mu}) \,,\label{op:1-} \\
  V_{x,\mu}^b(l) & = & \frac{1}{2}{\mathrm{tr}}(\tau^b\Phi^+_x
  U_{x,\mu}\ldots 
  U_{x+(l-1)\hat \mu,\mu}\Phi_{x+l\hat \mu})\, . 
\end{eqnarray}
Since already the masses of the lowest states in the $2^{++}$ channel are
relatively heavy we have not included Wilson loop contributions here.
Therefore, the wave functions of $2^{++}$ excited states have to be taken with
care.

In our procedure, finding the eigenfunctions $\tilde{\Psi}_i^{(n)}$ is
tantamount to determine an optimal source operator ${\mathcal{O}}^{(n)}$ for
the eigenstate $|{\bf \Psi}^{(n)}\rangle$ as a superposition of the original
operators ${\mathcal{O}}_i$,
\begin{equation}
  \label{eq:new_op}
  {\mathcal{O}}^{(n)}=\sum\limits_{i=1}^{N} \Psi_i^{(n)} {\mathcal{O}}_i \, ,
\end{equation}
where the coefficients $\Psi_i^{(n)}$ are related to the normalised solutions
$\tilde{\Psi}_i^{(n)}$ of the generalised eigenvalue equation
(\ref{eq:gen_eigen}) by $\Psi_i^{(n)}=L^{T~-1}_{ij}
\tilde{\Psi}_j^{(n)}/(\mu^{(n)}(t_0,t_0))^{1/2}$.  These coefficients define
the weight of the chosen operators contributing to ${\mathcal{O}}^{(n)}$ and
express the spatial extendedness of the eigenstate $|{\bf \Psi}^{(n)}\rangle$
under study.

In the $1^{--}$ and $2^{++}$ channels the index $i$ is used to label the size
$l$ of the corresponding Higgs string operator in lattice units.  In the case
of the $0^{++}$ channel the label $l=0$ refers to the operator $\rho_x^2$, and
for $l \geq 1$ we have to distinguish between the $l^{\mathrm{th}}$
contribution from Higgs strings and that from Wilson loops.
 
\begin{figure}[!htb]
    \begin{center}
      \epsfig{file=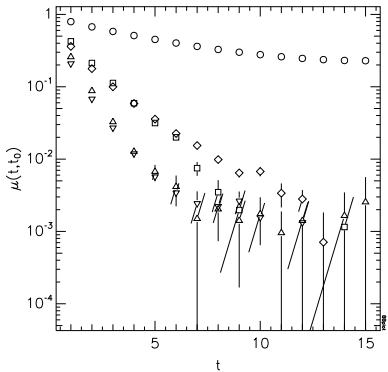,height=5cm,angle=0}
      \caption[]{Example of the exponential decay of $\mu^{(n)}(t,t_0)$
        for the five lowest states}
      \vspace{7mm}   
      \label{fig:logplot}
    \end{center}
\end{figure}

We have performed simulations using the update algorithms as described in our
previous works \cite{wirNP97}. In the error analysis of the wavefunctions and
masses we have used the jackknife technique.  As an example of the outcome of
the analysis we show in Fig.~\ref{fig:logplot}
\begin{figure}[!htb]
    \begin{center}
      \epsfig{file=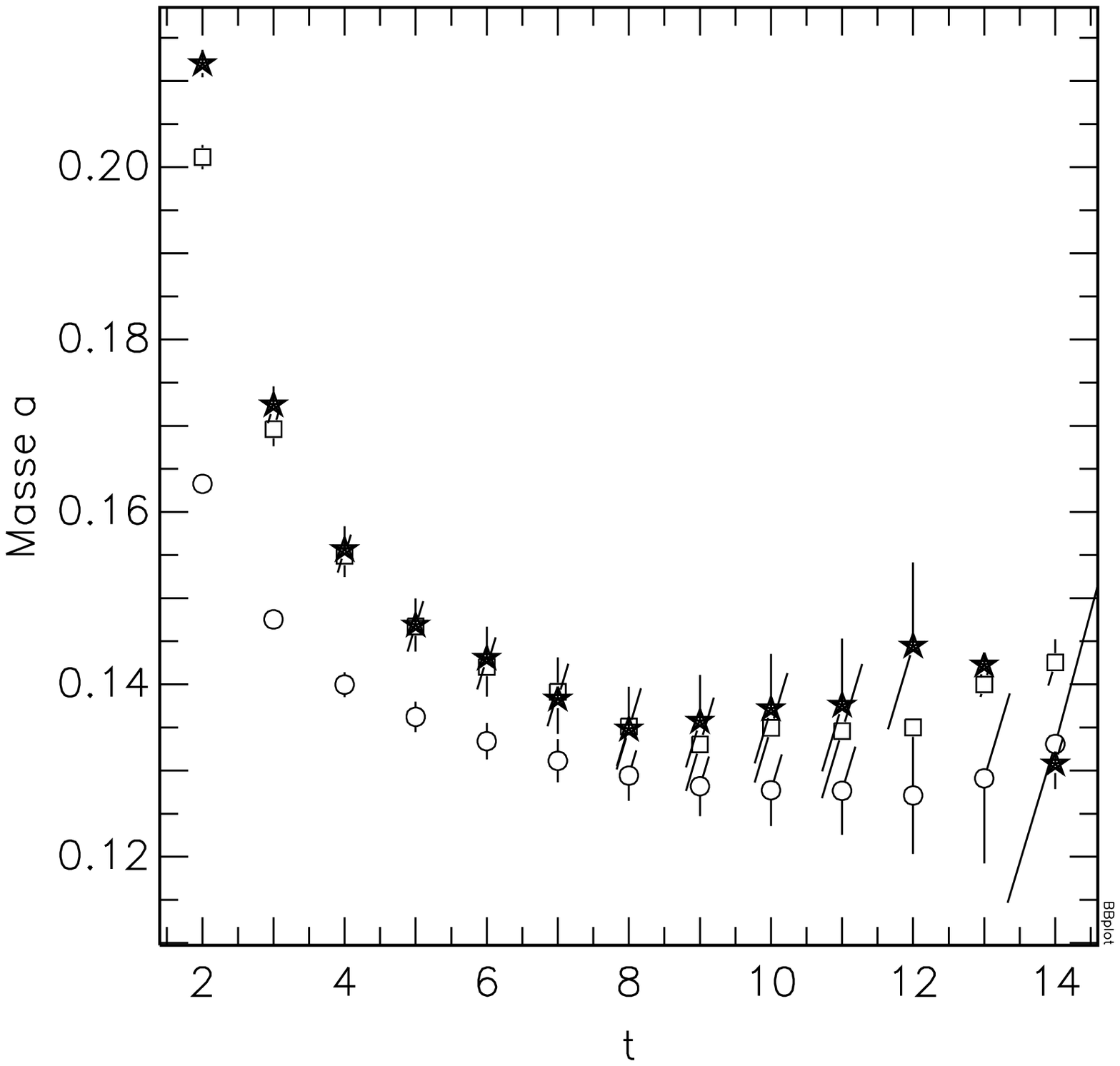,height=5cm,angle=0} 
      \caption{$0^{++}$ ground state effective mass obtained from cross
        correlations $\left( \kreisl \right)$, compared to results solely from
        $S_{x,1}+S_{x,2}$~$\left( \bbox \right)$ and $\rho_x^2$~$\left( \stern
        \right)$.}
      \label{fig:logplot1}
    \end{center}
\end{figure}
the exponential decay of the diagonal elements of the correlation matrix
$\mu^{(n)}(t,t_0)$ for the first five states ($n=1,\ldots,5$) in the $0^{++}$
channel as they are found in the symmetric phase (at $\beta_G=12,
\beta_H=0.3434$) for a Higgs mass of $M_H^*=70$ GeV.  The mass estimates are
obtained by fitting $\mu^{(n)}(t,t_0)$, throughout a plateau region of the
effective mass, to a hyperbolic cosine form, taking into account the estimated
errors of the correlation functions.  The improvement achieved by using cross
correlations is demonstrated in Fig.~\ref{fig:logplot1}.  As expected, the
cross correlation method leads to lower mass values.

The measurements for a Higgs mass $M_H^*=70$ GeV ($\lambda_3/g_3^2\approx
0.09570$) were performed below but near to the endpoint of the phase
transition.  At both sides, in the symmetric and in the Higgs phase, we used a
respective line of constant physics corresponding to
$m_3^2(g_3^2)/g_4^2\approx 0.00023 $ in the symmetric and
$m_3^2(g_3^2)/g_4^2\approx-0.043$ in the Higgs phase.  This corresponds to
$T\approx 1.02 T_c$ and $T\approx 0.98 T_c$, respectively
\cite{generic,wirNP97}.  The gauge coupling $\beta_G$ has been varied from 8
to 16 in order to examine the extrapolation to the continuum.  The choice of
the actual simulation parameters as near as possible to the phase transition
line was dictated by the necessity to avoid tunnelling between the phases.
Besides the used lattice size of $30^3$ we have additionally investigated the
wave functions in the symmetric phase for the $1^{--}$ and $2^{++}$ channels
on a $50^3$ lattice which permitted to increase the range of $l$ in the
operator basis.  On a lattice of given size, the spatial extension of
operators is restricted to half of the lattice size.

At the larger Higgs mass of $M_H^*=100$ GeV ($\lambda_3/g_3^2\approx 0.1953$),
above the endpoint of the phase transition \cite{endpointh,endpoint}, we have
studied the spectral change with decreasing temperature (increasing
$\beta_H$), from the ''symmetric'' to the ''Higgs'' side of the crossover
line, only at one fixed $\beta_G=12$.

The accumulated statistics of our measurements in the symmetric and Higgs
phases as well as in the crossover region for each pair of $\beta_G$ and
$\beta_H$ values is summarised in Table \ref{tab:statistics}.
\begin{table*}[!thb]
  \caption[]{Used statistics per ($\beta_G$, $\beta_H$) in the measurements}
  \label{tab:statistics}
  \centering
    \begin{tabular}{|l|r|r|c|c|r|}\hline
      &&&&&\\[-2 ex]
      Phase    & Channel       &$M_H^*$& $L^3$ & $\beta_G$ & independent    \\
      
      &               &   GeV &       &           & configurations \\[0.5 ex]
      \hline 
      &&&&&\\[-2 ex]
      symmetric&$1^{--},2^{++}$& 70    &$50^3$ & 8,12,16   &  20000 \\[0.5 ex]
      \cline{2-6}\cline{5-6}
      &&&&&\\[-2 ex]
      & $0^{++},1^{--},2^{++}$ & 70  & $30^3$  & 8,12,16 &  12000 \\ [0.5 ex]\hline
      &&&&&\\[-2 ex]
      Higgs &  $0^{++}$,$1^{--},2^{++}$&70  & $30^3$ & 8,12,16 & 12000 \\ [0.5 ex]\hline 
      &&&&&\\[-2 ex]
      crossover &  $0^{++}$,$1^{--},2^{++}$ & 100 &  $30^3$ & 12
      & 4000 \\[0.5 ex] \hline
    \end{tabular}
\end{table*}
By measuring the autocorrelations for the operators of interest we determined
the frequency of measurements during the updating in order to retain only
independent configurations.


\section{Bound States in the Symmetric Phase at {\boldmath $M_H^*=70$}  GeV}

Using the cross correlation technique we were able to obtain the wave function
squared corresponding to the optimised operator for each individual state in
the spectrum. Being functions of a physical distance, the squared wave
functions are shown immediately {\it vs.} $l~a~g_3^2~$ in order to overlay
data from measurements at various gauge couplings (lattice spacings) taken
along a line of constant physics.

\begin{figure}[!htb]
  \begin{minipage}{8.8cm}
    \begin{center}
      \epsfig{file=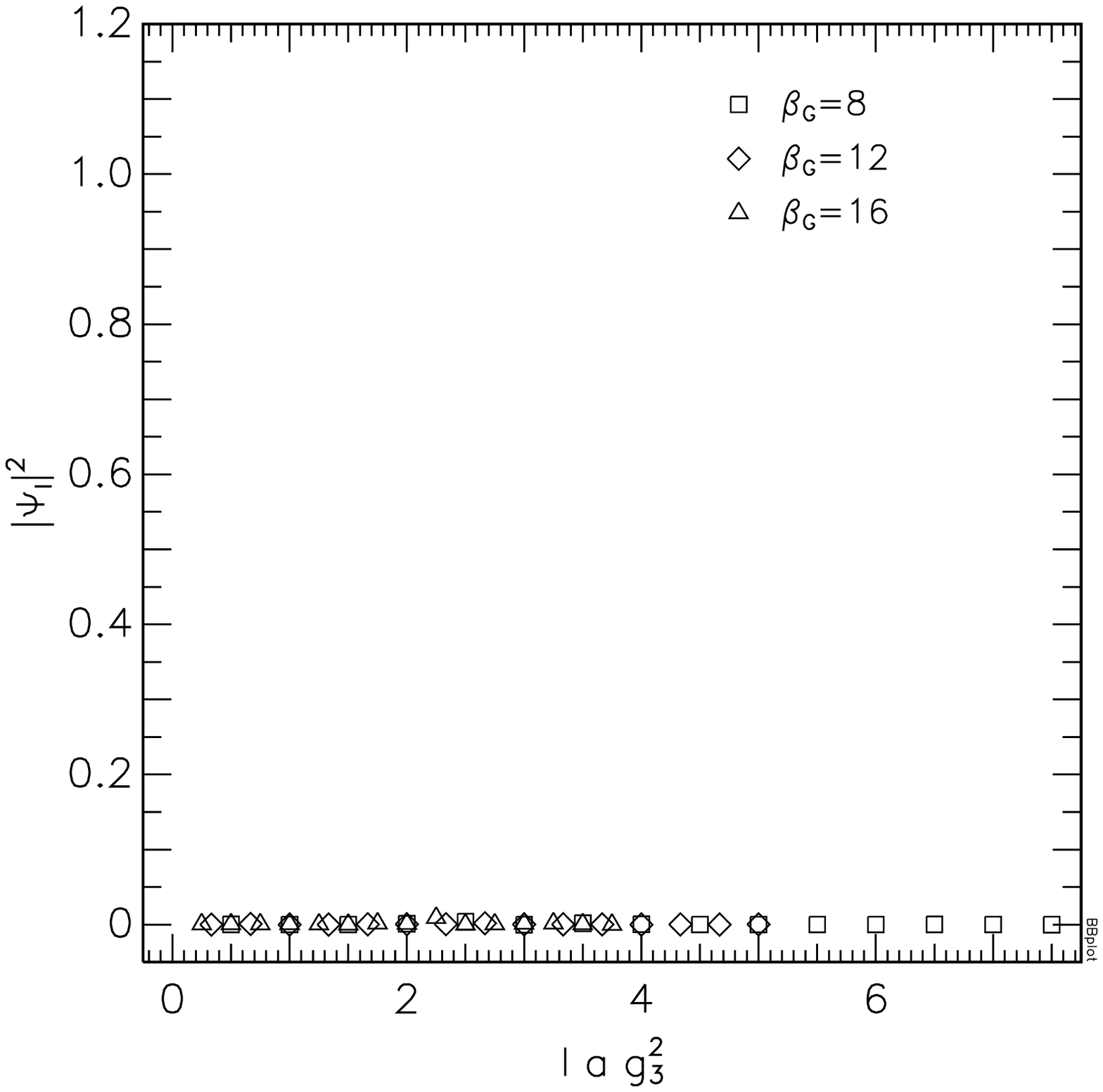,width=4.3cm,height=4.3cm,angle=0}
      \epsfig{file=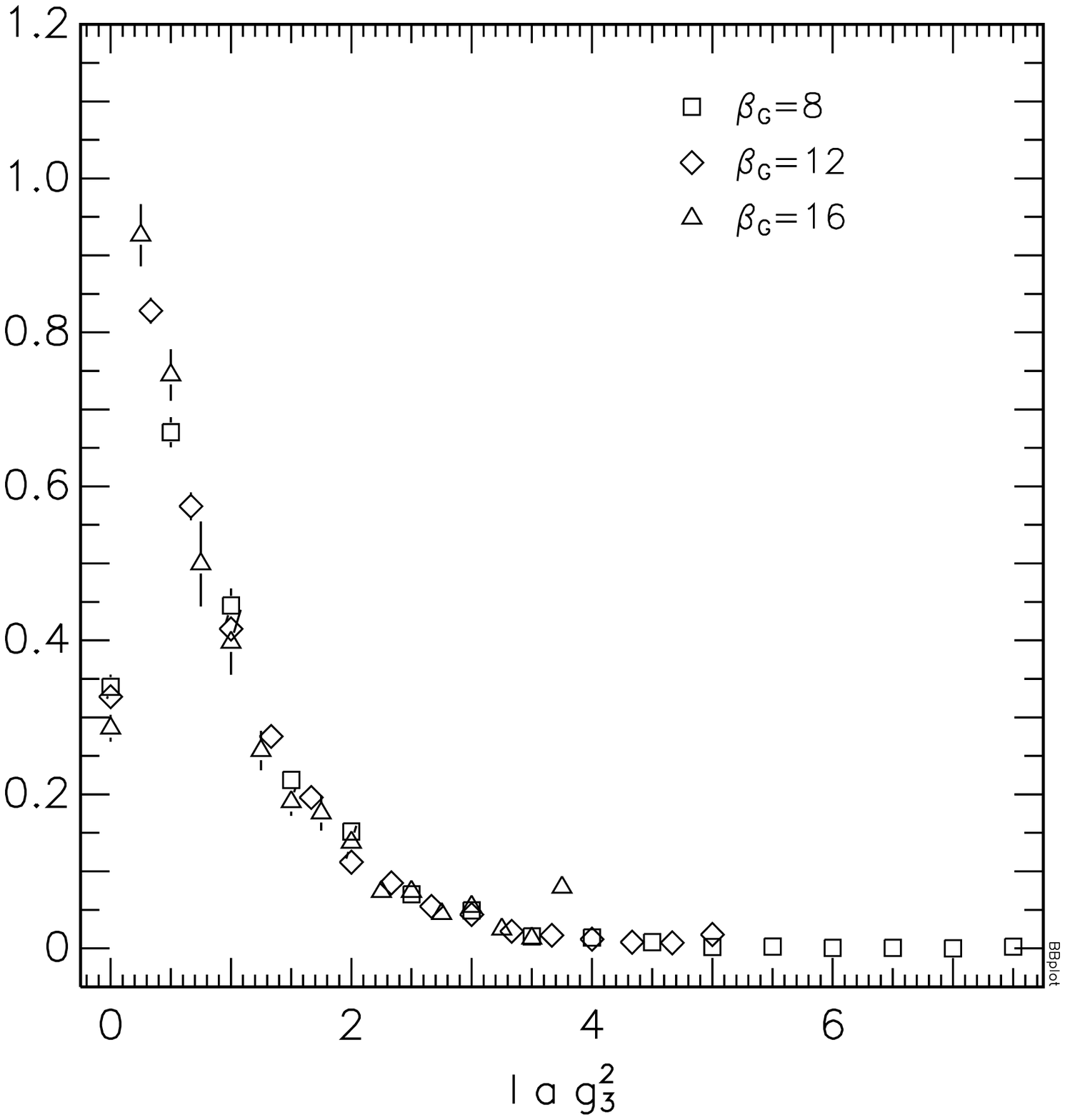,width=4.1cm,height=4.3cm,angle=0}
      \caption[]{Squared wave function of the ground state in the 
        $0^{++}$ channel,
        measured on a $30^3$ lattice in the
        symmetric phase; 
        left: $W_{x,1,2}(l)+W_{x,2,1}(l)$,
        right: $S_{x,1}(l)+S_{x,2}(l)$; 
        $l=0,\ldots, 15$}
      \label{fig:kontwave_10+}
    \end{center}
  \end{minipage}
\end{figure}

\begin{figure}[!htb]
  \begin{minipage}{8.8cm}
    \begin{center}
      \epsfig{file=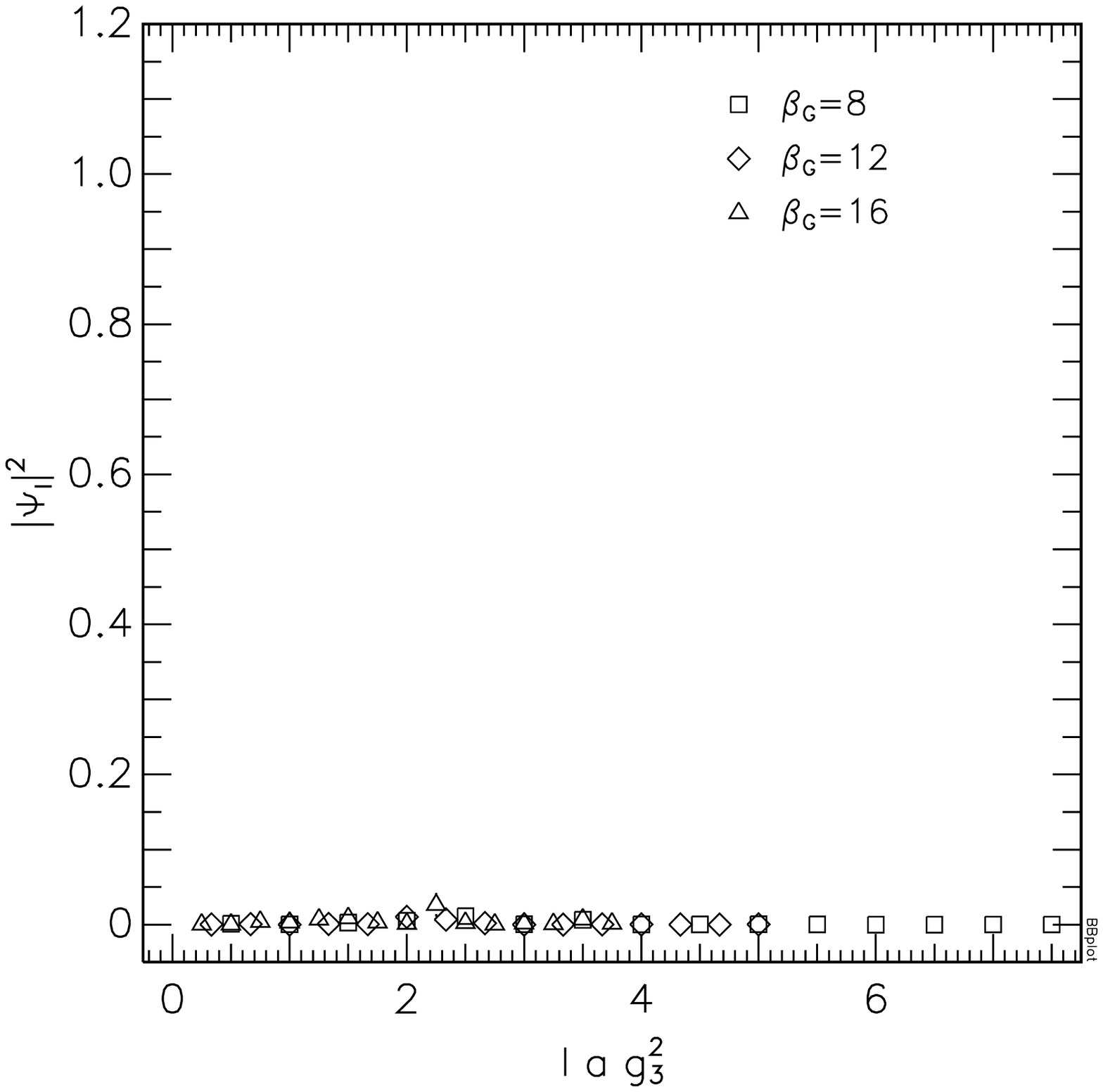,width=4.3cm,height=4.3cm,angle=0}
      \epsfig{file=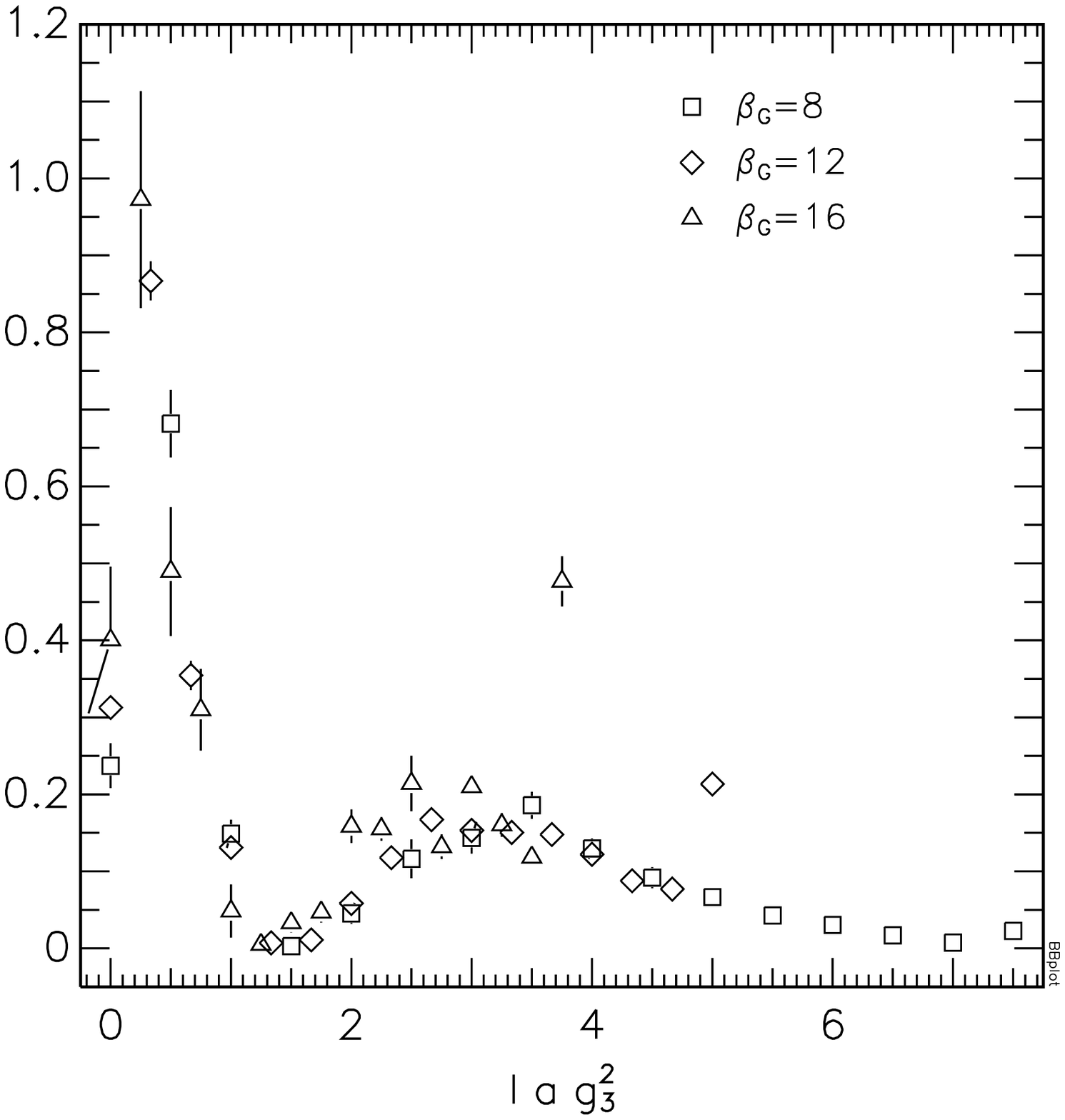,width=4.1cm,height=4.3cm,angle=0}
      \caption[]{Same as Fig.~\ref{fig:kontwave_10+} for the 
      first excited state}
      \label{fig:kontwave_20+}
    \end{center}
  \end{minipage}
\end{figure}

\begin{figure}[!htb]
  \begin{minipage}{8.8cm}
    \begin{center}
      \epsfig{file=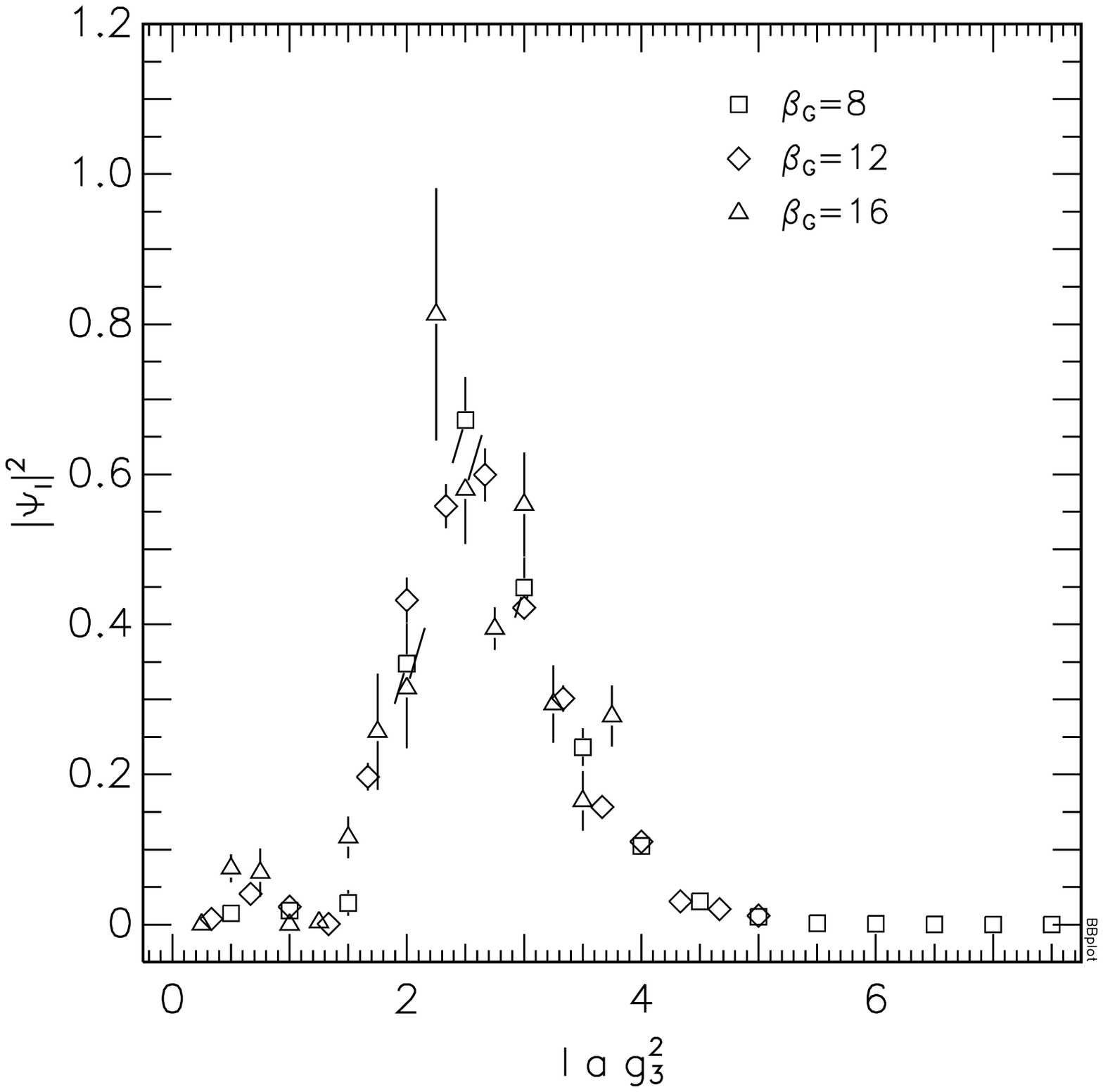,width=4.3cm,height=4.3cm,angle=0}
      \epsfig{file=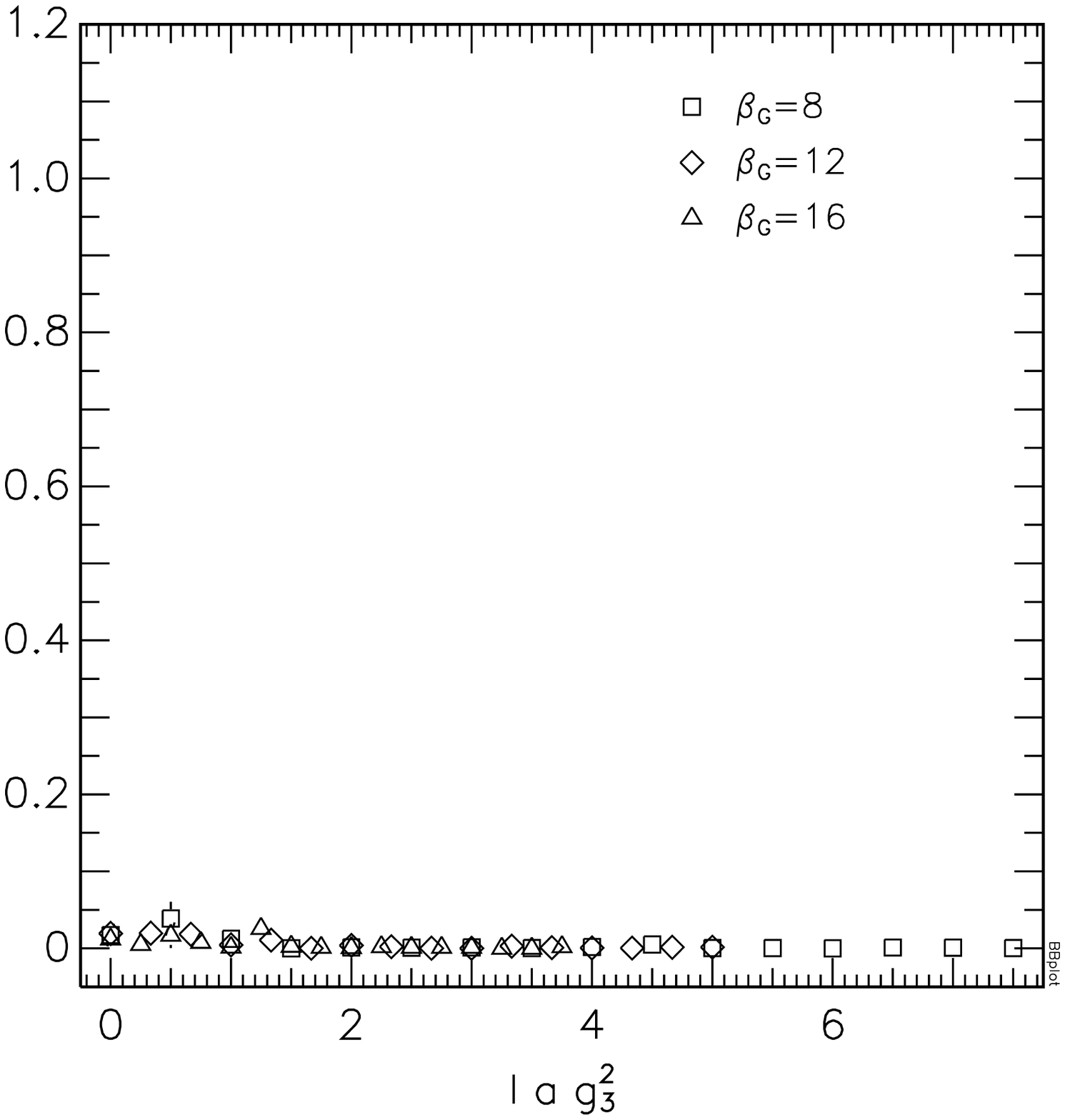,width=4.1cm,height=4.3cm,angle=0}
      \caption[]{Same as Fig.~\ref{fig:kontwave_10+} for the
      second excited state }
      \label{fig:kontwave_30+}
    \end{center}
  \end{minipage}
\end{figure}

The results for the $0^{++}$ channel are collected in
Figs.~\ref{fig:kontwave_10+}-\ref{fig:kontwave_30+} for the squared
wave functions. The contributions from the Higgs string and Wilson loop
operators are shown separately in order to identify clearly Higgs and $W$-ball
excitations.  We observe no mixing of these two operator types in the Higgs
ground state and the first excitation.  The second excited state in this
channel consists of a pure excitation of gauge degrees of freedom (d.o.f.) and
can be identified with a $W$-ball in analogy with the glueballs of pure
$SU(2)$. Our results on this decoupling confirm the observations in
\cite{philipsen} made at a much lighter Higgs mass in the symmetric phase near
to the strongly first order phase transition.
  
\begin{figure*}[!thb]  
  \begin{minipage}{16cm}
  \centering 
    \epsfig{file=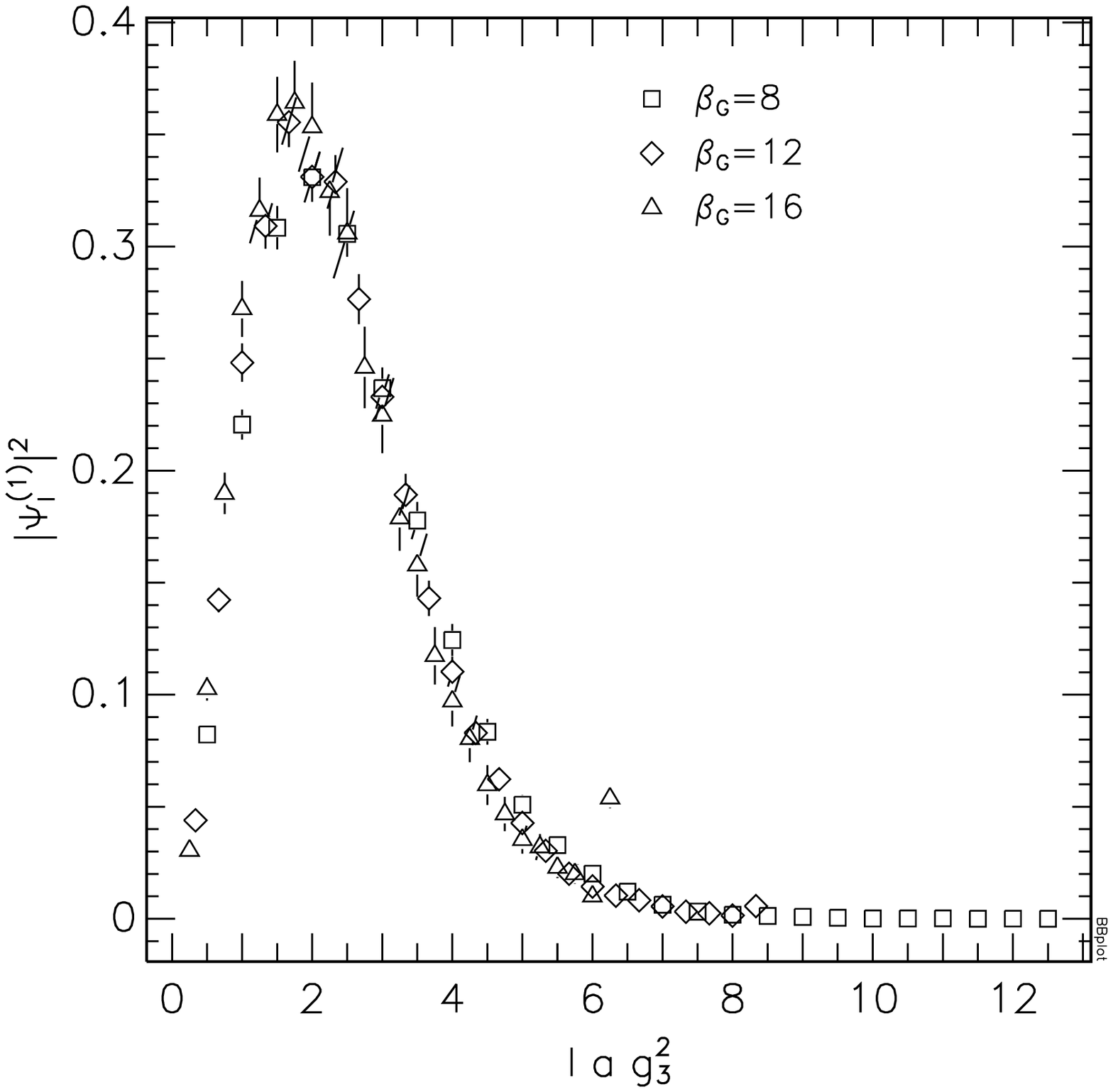,width=5cm,height=5cm,angle=0} 
    \epsfig{file=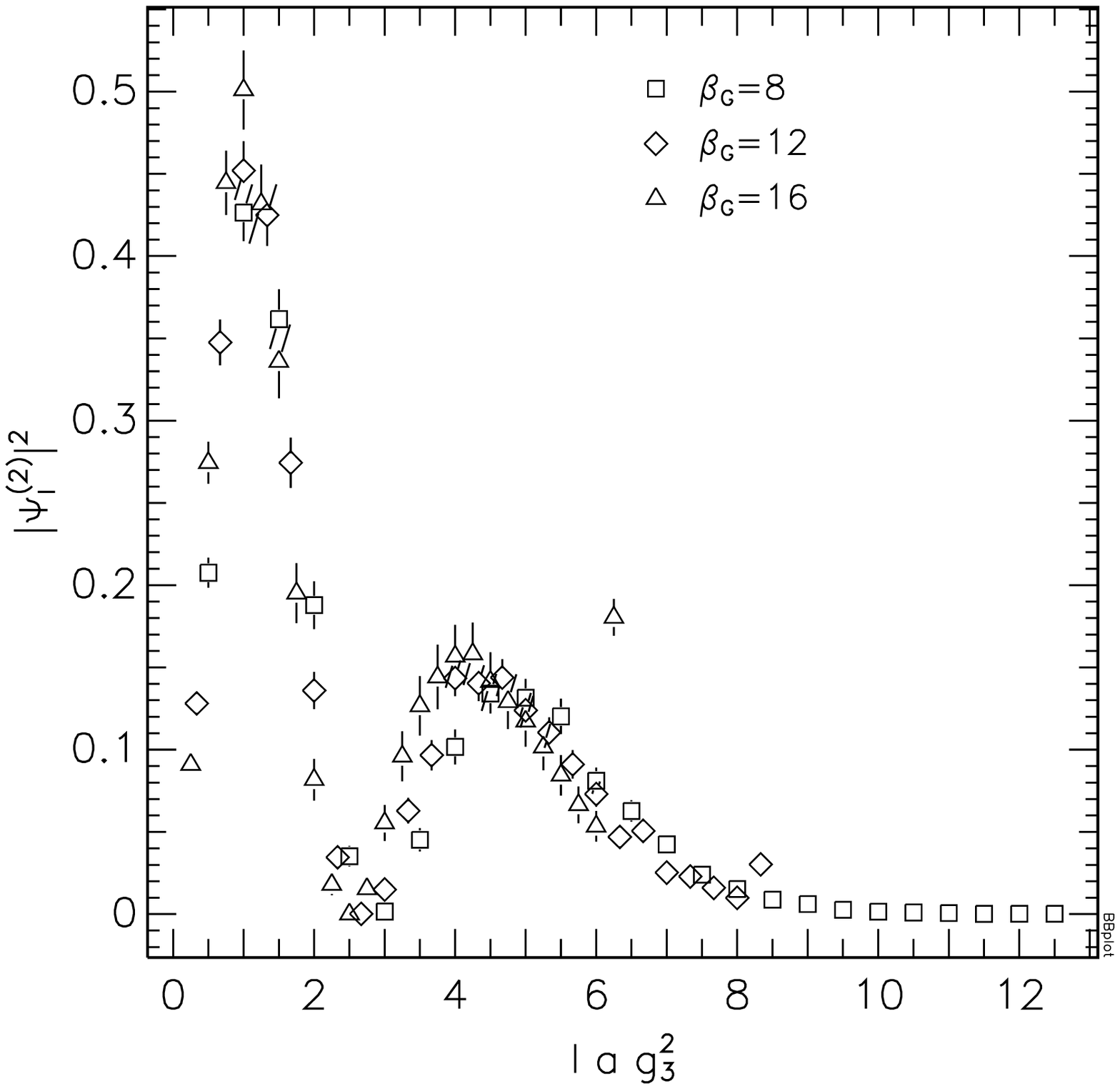,width=5cm,height=5cm,angle=0}
    \epsfig{file=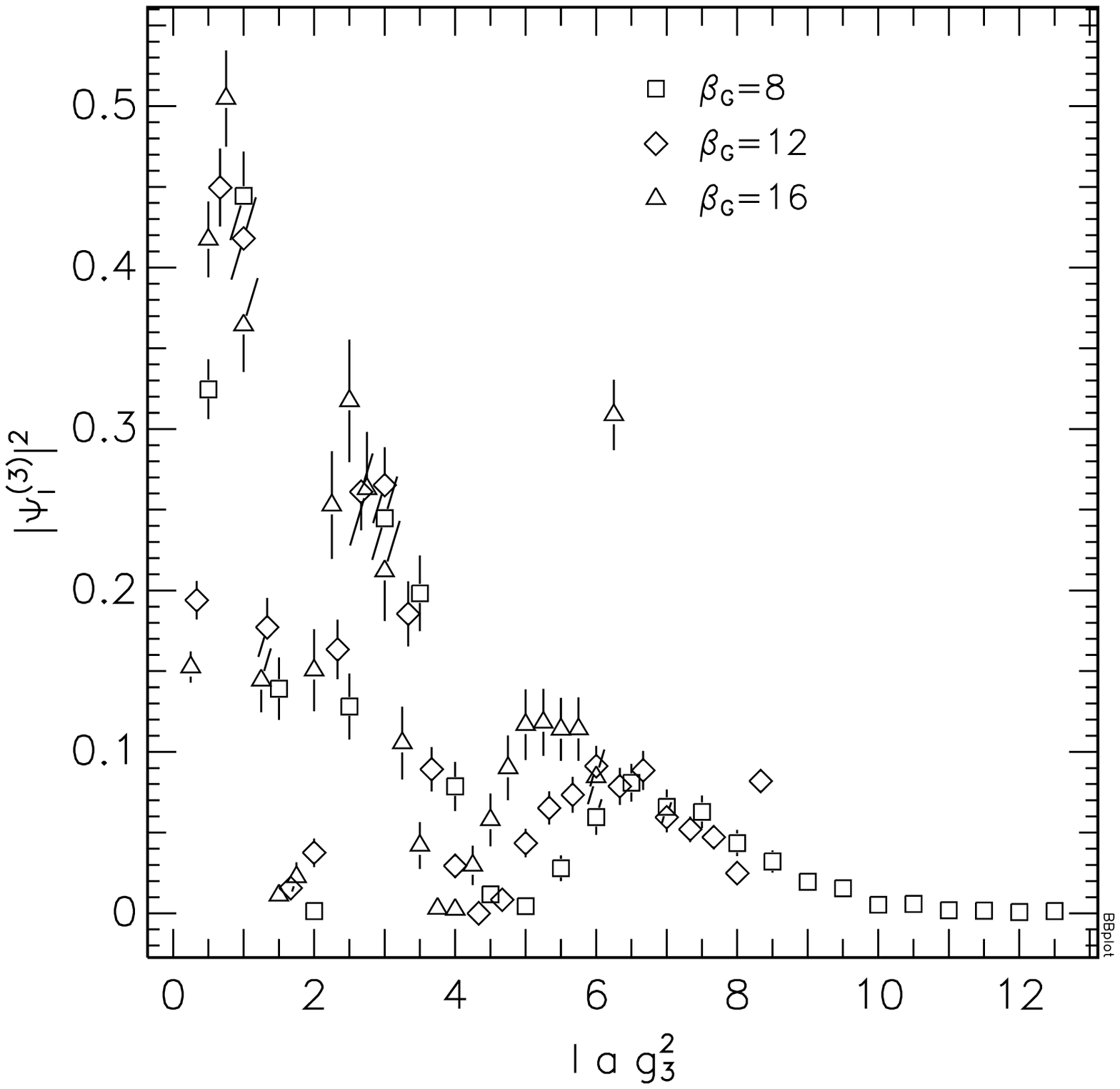,width=5cm,height=5cm,angle=0}
      \caption[]{Squared wave function in the $1^{--}$ channel, 
        measured on a
        $50^3$  lattice in the symmetric phase; 
        ground state and two excited states}
      \label{fig:kontwave_1--}
  \end{minipage}
\end{figure*}

\begin{figure}[!htb]  
  \begin{minipage}{8.8cm}
    \begin{center}
      \epsfig{file=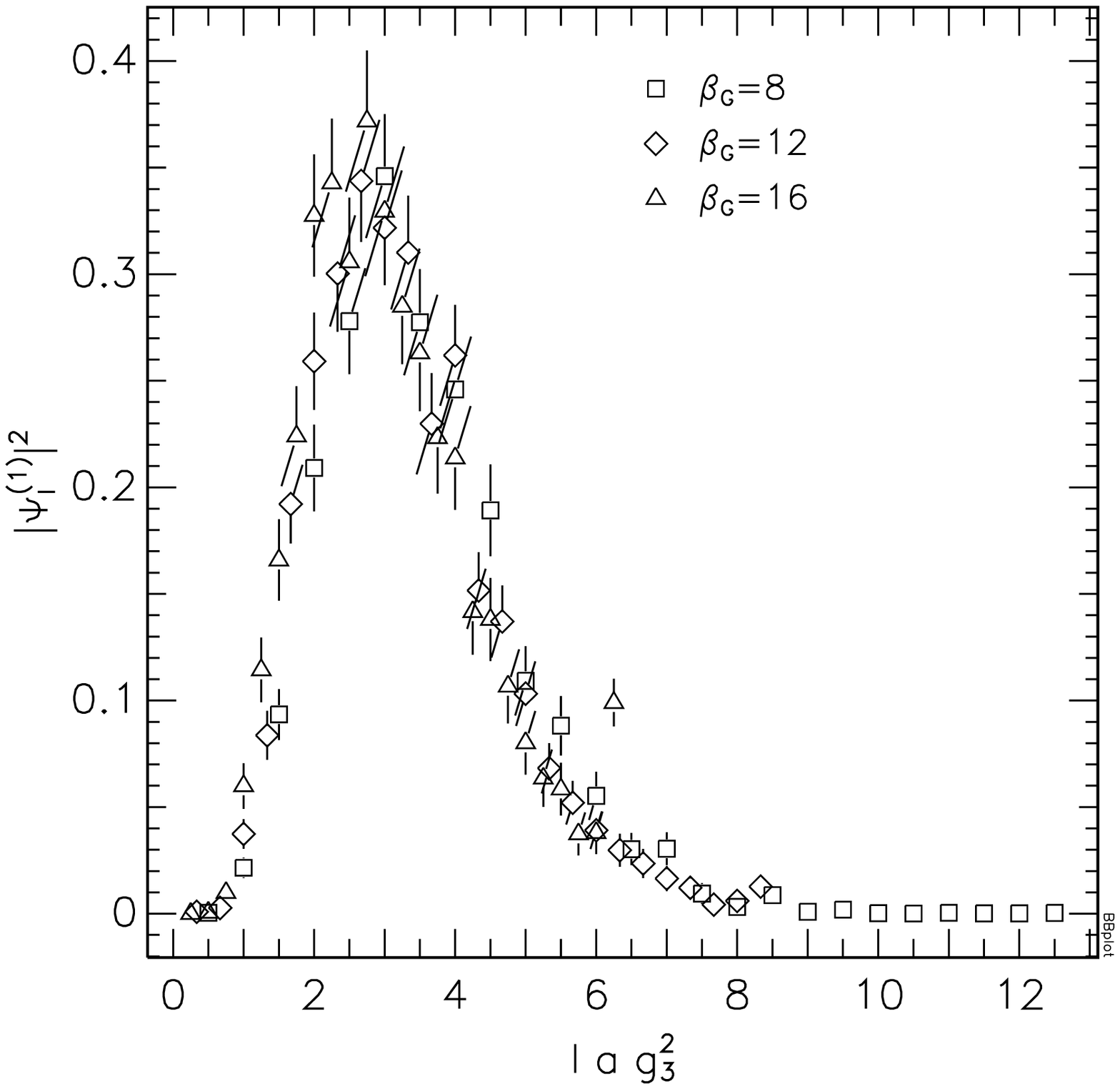,width=4.2cm,height=4.3cm,angle=0}
      \epsfig{file=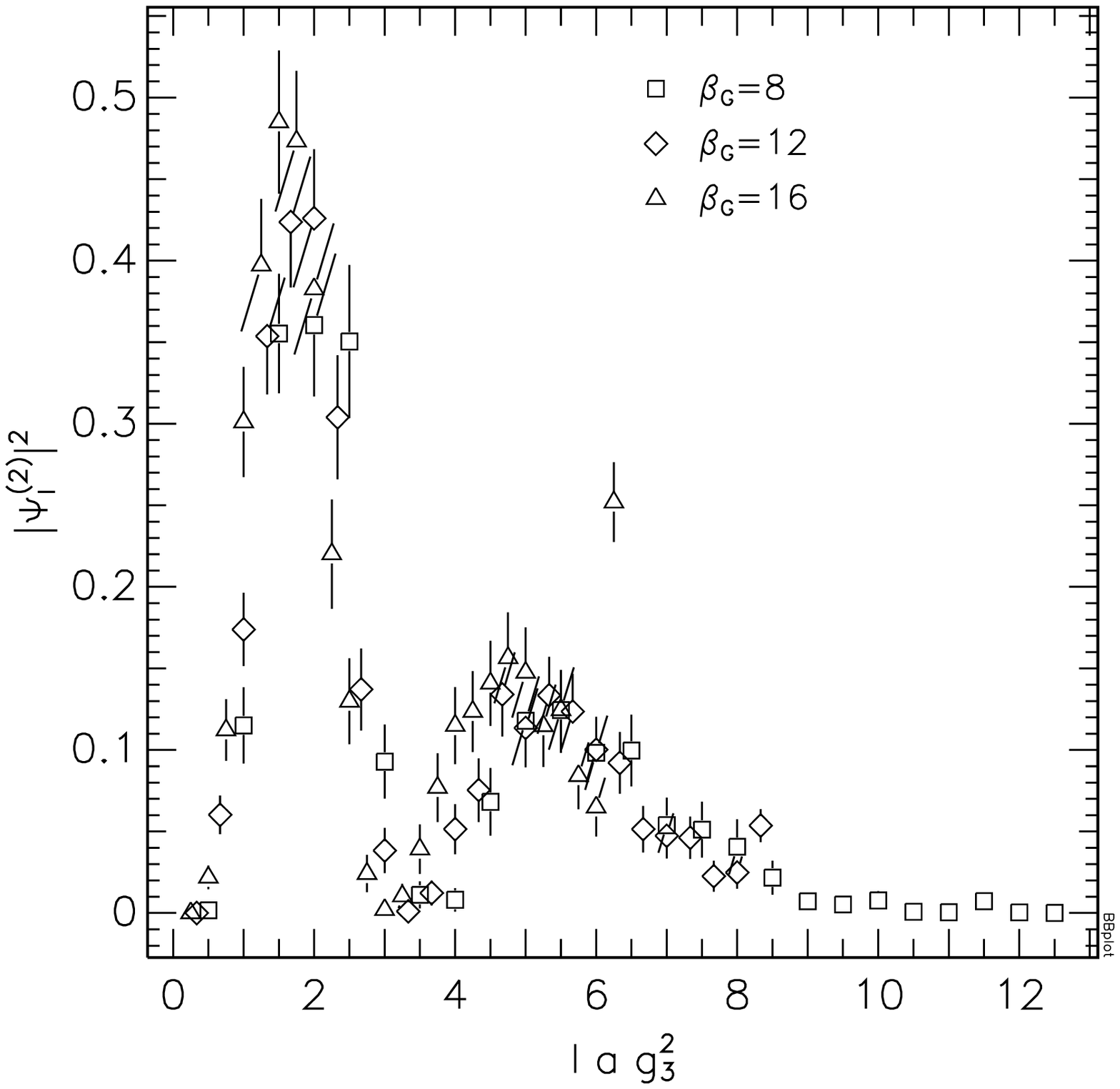,width=4.2cm,height=4.3cm,angle=0}
      \caption[]{Squared wave function in the $2^{++}$ channel,
        measured on a
        $50^3$  lattice in the symmetric phase; 
        ground state and first excited state}
      \label{fig:kontwave_2++}
    \end{center}
  \end{minipage}
\end{figure}

The squared wave functions for ground and excited states in the $1^{--}$ and
$2^{++}$ channels are presented in Fig.~\ref{fig:kontwave_1--} and
\ref{fig:kontwave_2++}.

At this point it is due to come back to the issues of completeness (of the
operator set) and orthogonality (of the states $\Psi_i^{(n)}$). We have
already noticed that the eigenvectors of (\ref{eq:gen_eigen_first}) become
orthogonal only in the limit of a complete operator set, and that the scalar
product between different states should be used to examine the degree of
completeness. This test shows what are the difficulties in practice.  We have
studied the scalar product between the ground state and the first excitation
as a function of the maximal length $l_{max}$ of the operator set
$S_{x,\mu}(l)$ by gradually clipping the set of operators from $l_{max}$=25
(on the $50^3$ lattice) down to zero. One would expect that this function
tends to zero for $l_{max}\to\infty$. Actually the scalar product reaches a
minimum at some finite $l_{max}$ and starts to grow at bigger maximal length
due to statistical and numerical errors in the case of larger operators and
matrices.

Inspecting the wave function of the ground state in the $1^{--}$ channel we
observe that for very large operators the contribution to this state vanishes.
This is true to a high accuracy for the largest physical volume that we have
considered, at $\beta_G=8$. If we chose higher $\beta_G$ (smaller volumes)
along the line of constant physics (in order to explore the approach to the
continuum) the operator set in use with fixed $l_{max}=L/2$ does not describe
anymore the whole wave function. There are fake contributions accumulating in
the contribution of the longest operator in the set which, in principle,
belong to more extended operators which are not included.  A similar behaviour
can be observed for the excited states if one compares the wave functions
varying the operator set.  For the second excited state shown in
Fig.~\ref{fig:kontwave_1--} it is clearly seen that in the case of
$\beta_G$=16 (small physical volume) this state is not satisfactorily
described by the limited operator basis.  An optimal continuum limit for the
wave function would require to choose the same maximal operator length in
physical units, {\it i.e.} doubling the length in lattice units when one goes
over from $\beta_G=8$ to $\beta_G=16$. This is, however, difficult to
accomplish with restricted resources, not only because of the lattice size.
Larger cross correlation matrices would additionally require larger statistics
to get stable results from the diagonalisation procedure.

Considering only wave function data obtained at fixed $\beta_G$ (tolerating a
non-perfect overlay of data from different $a$ in a physical scale) the
presented figures illustrate perfectly that the number of zeroes of the wave
functions corresponds to the level of excitation (knot rule).

Due to the mentioned difficulties, it becomes more and more difficult to
extract the masses of higher excitations, in particular if already the ground
state is very heavy and if the lattice becomes dangerously small (when we
attempt to study the continuum limit) compared to the physical extension of
the wave function corresponding to the state under discussion.

For the ground states we are now able to reexamine our previous spectrum
investigations \cite{wirNP97} where we measured the mass without use of the
cross correlation technique. In the $0^{++}$ channel the agreement is
satisfactory since the ground state wave function is well dominated by the
shortest operators like $(1/2){\mathrm{tr}}(\Phi^+_x U_{x,\mu}\Phi_{x+\hat
  \mu})$.  Mass measurements in the $2^{++}$ channel and for the $W$-ball
states have proved to be difficult in our previous studies, and we did not
report on that in Ref. \cite{wirNP97}. From the present analysis we recognise
that this difficulty is related to the large spatial extendedness of these
states.

Using the cross correlation technique we have found in the symmetric phase the
ground state mass and the first excitations. We restricted ourselves to an
identification only of those masses with a still reasonable plateau behaviour
and a not too large mass value in lattice units.  Notice that the operators
are classified in the angular momentum $J$ only modulo $4$; therefore we
cannot exclude that states with higher masses have got contributions from
states with higher $J$.

\begin{figure*}[!thb]
  \begin{minipage}{16cm}
  \centering 
    \epsfig{file=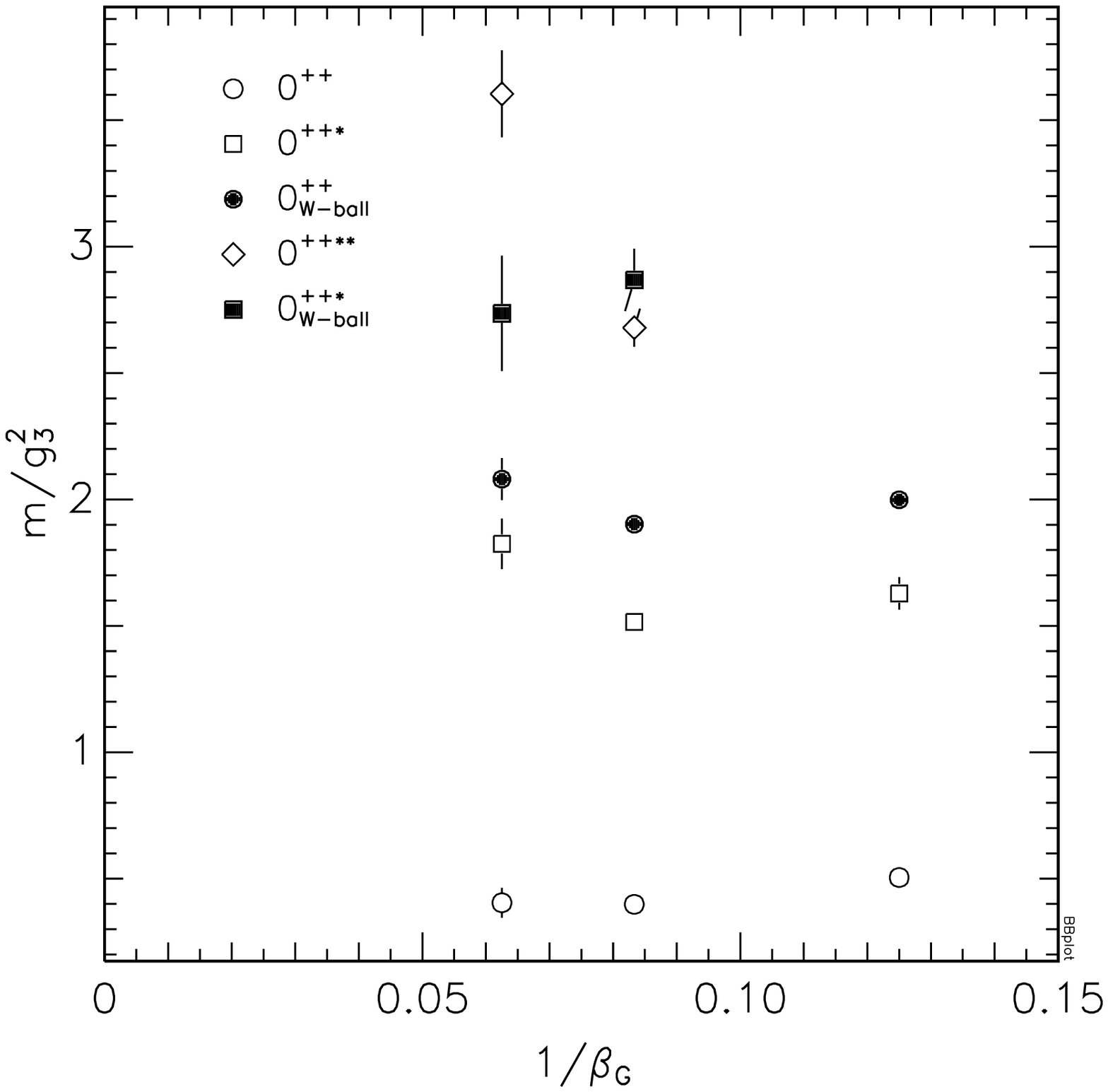,width=5cm,height=5cm,angle=0} 
    \epsfig{file=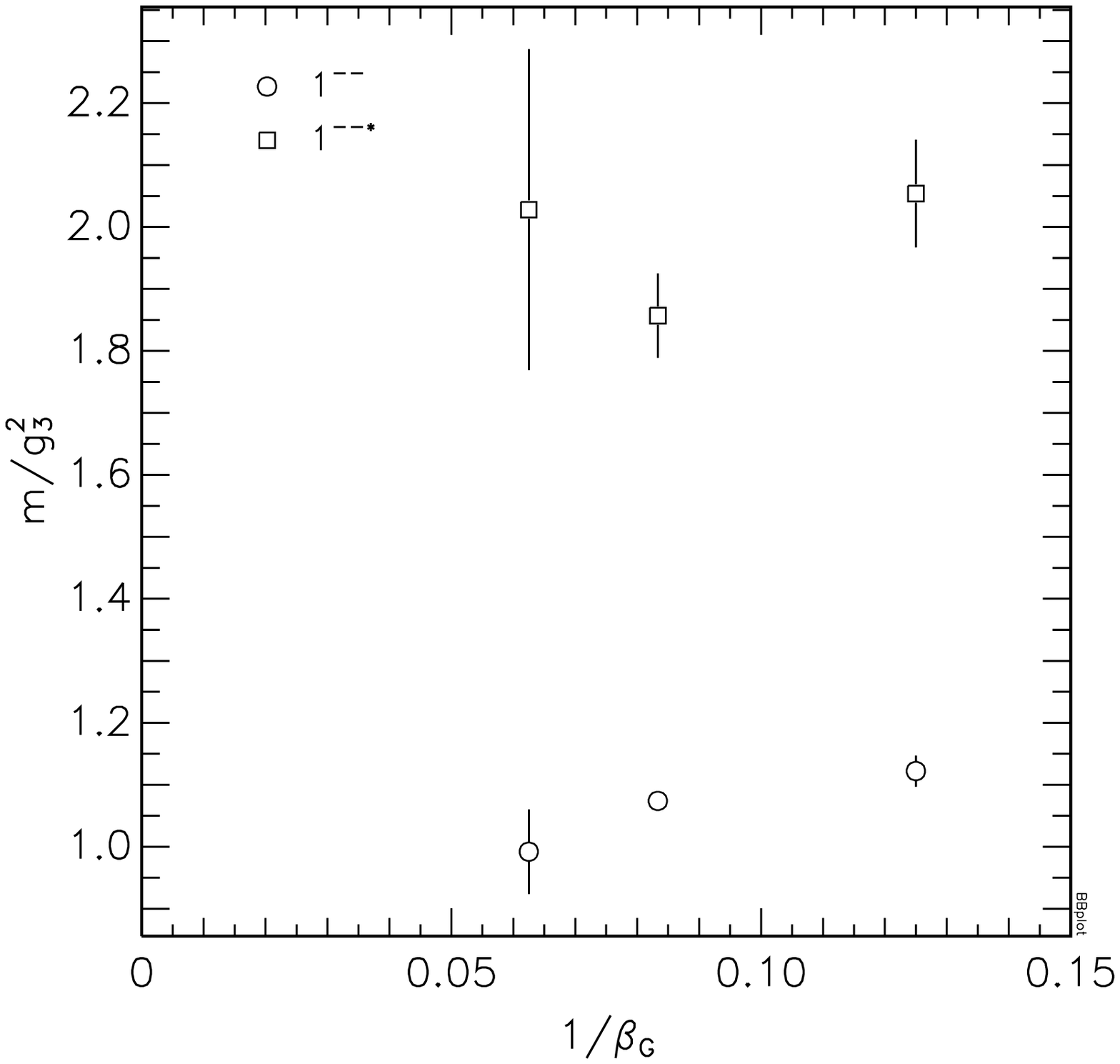,width=5cm,height=5cm,angle=0}
    \epsfig{file=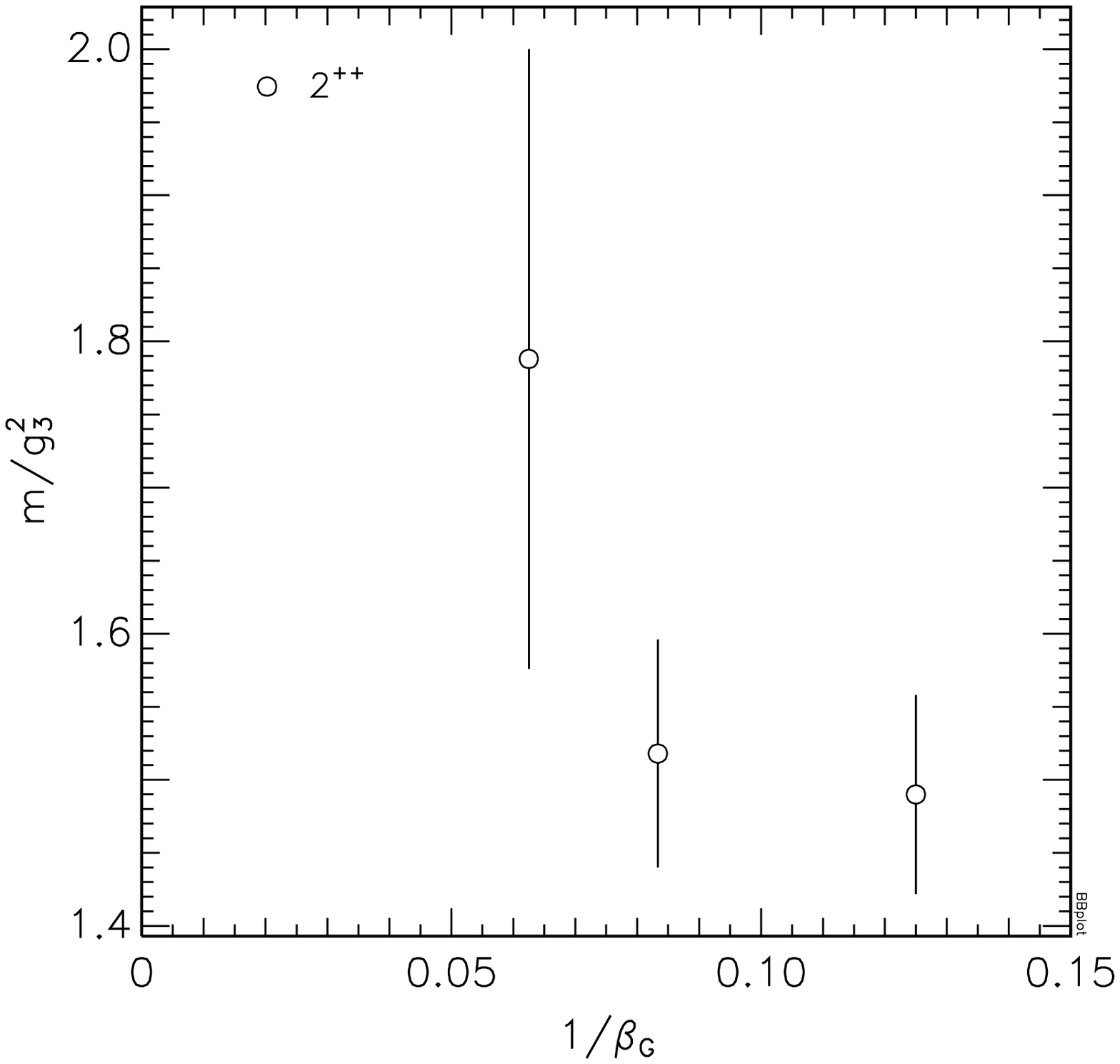,width=5cm,height=5cm,angle=0}
    \caption[]{Mass spectrum in the symmetric phase as function of 
     $1/\beta_G$}
    \label{fig:kont_spectrum}
  \end{minipage}
\end{figure*}

The masses of the first excitations in the symmetric phase (for $2^{++}$ only
of the lowest state) are presented in Fig.~\ref{fig:kont_spectrum} and the
fitted masses are collected in
Tables~\ref{tab:masskont0+}-\ref{tab:masskont1--2++} of Appendix~\ref{appB}.
We tried to illustrate whether an approach to the continuum limit is already
indicated by the present measurements. The big errors of the excited state in
the $1^{--}$ channel and, even more, of the ground state in the $2^{++}$
channel show where the problems are.

Concluding this Section we would like to stress that qualitatively there is no
difference in the spectrum on the high temperature side of the phase
transition for light or relatively heavy Higgs masses as long as the phase
transition persists.  The gauge d.o.f. decouple from the Higgs excitations and
form $W$-balls similar to confined $3D$ pure $SU(2)$ gauge theory
\cite{tepergb}.


\section{The Spectrum in the Higgs Phase near to the Endpoint of the
Transition}

On the Higgs side of the phase transition pure gauge matter ($W$-ball)
excitations are not expected to be present in the spectrum.  This hypothesis
has been numerically verified at light Higgs masses \cite{philipsen}.  Now, at
$M_H^*=70$, we observe in the Higgs phase near to the transition the
following.  In the $1^{--}$ and $2^{++}$ channels the spectrum looks similar
to that in the symmetric phase.  This was expected from earlier studies for
the $1^{--}$ channel.  Therefore the squared wave functions are not shown
explicitly.

Note, that in the $2^{++}$ channel we have not used operators with pure gauge
degrees of freedom.  Therefore, we cannot observe the expected difference for
$2^{++}$ $W$-balls.\footnote{Results for the $W$-balls with $J=2$ can be found
  in \cite{philipsen1}.}

\begin{figure}[!htb] 
  \begin{minipage}{8.8cm}
    \begin{center}
      \epsfig{file=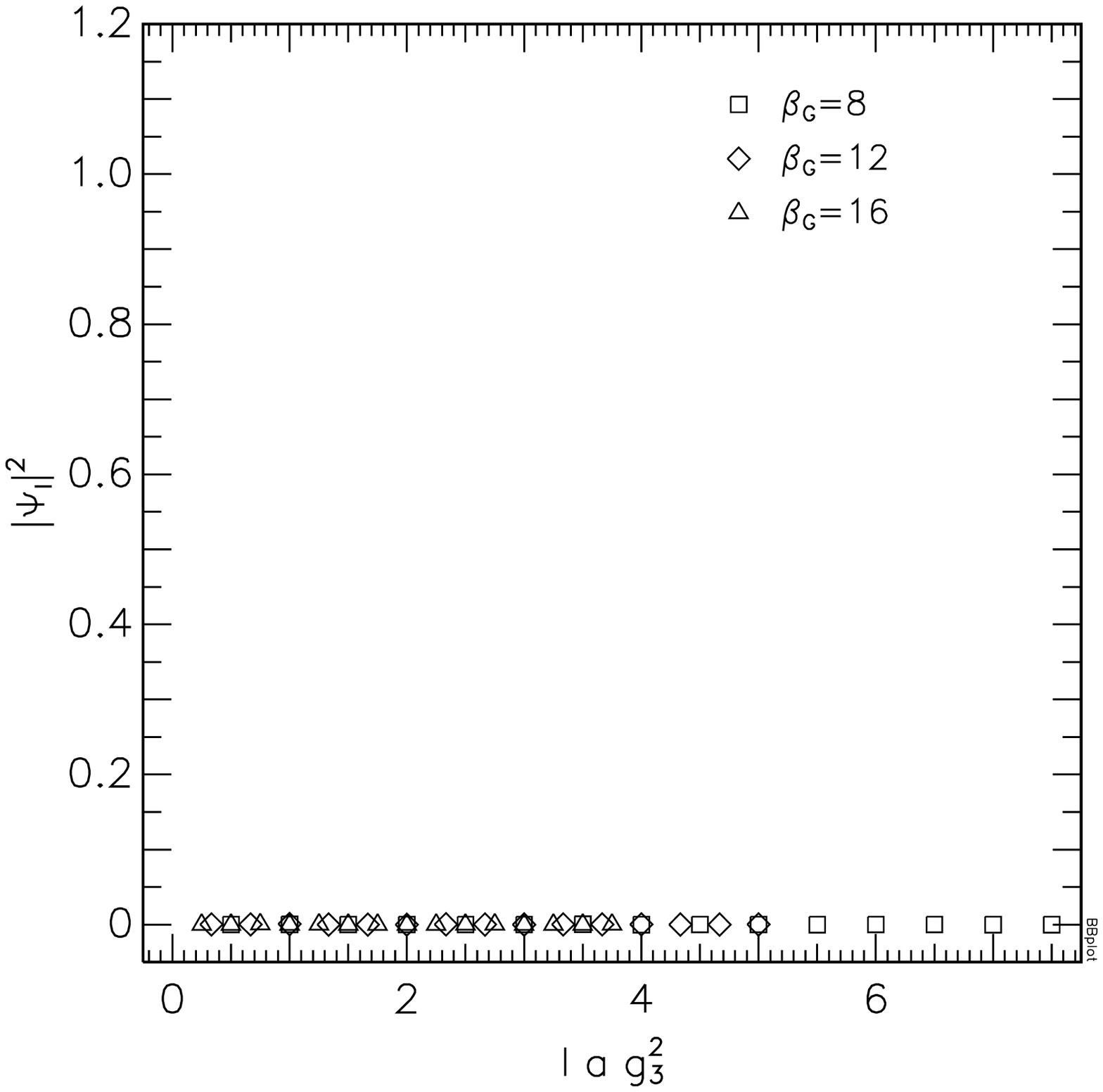,width=4.3cm,height=4.3cm,angle=0}
      \epsfig{file=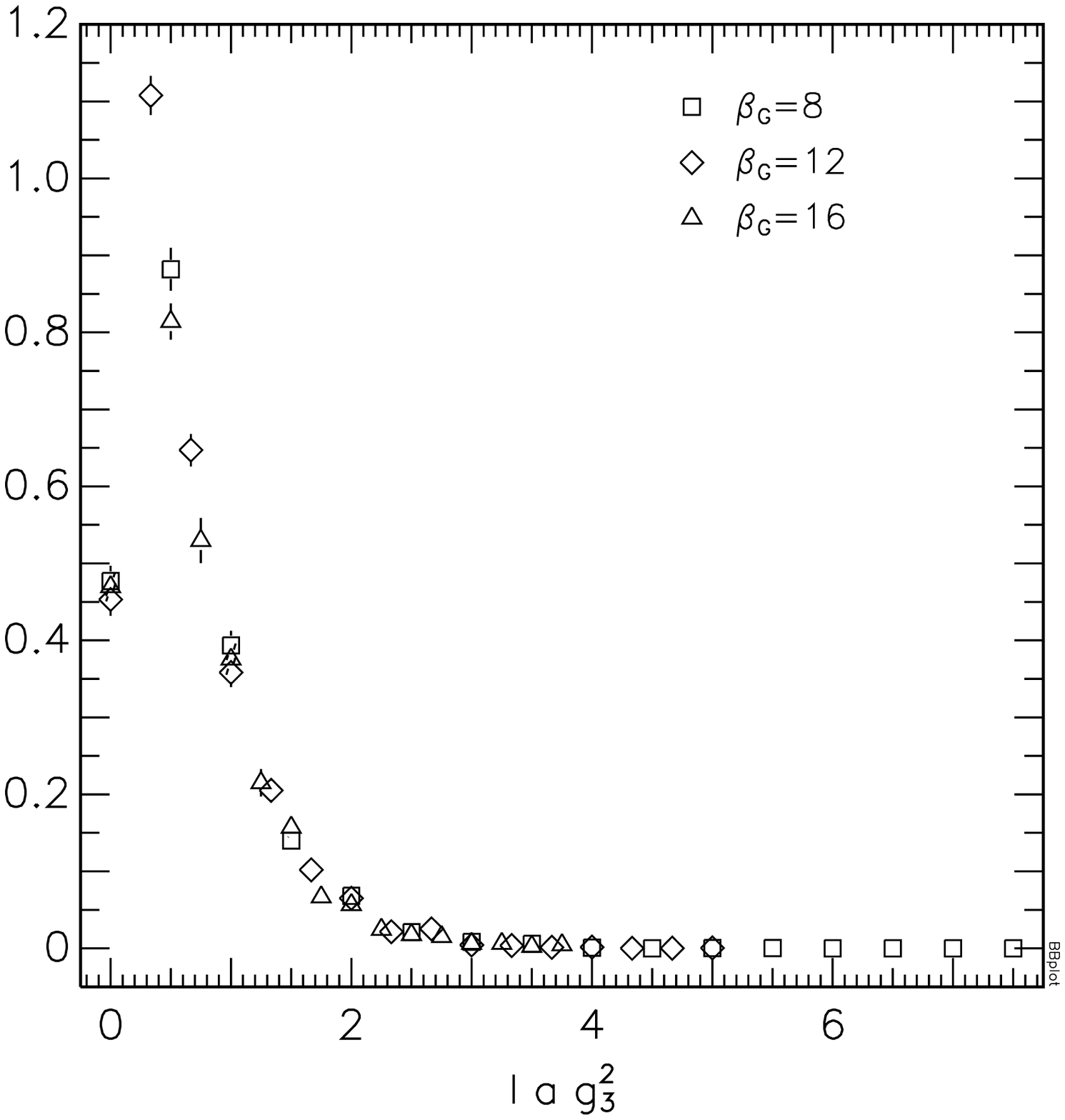,width=4.1cm,height=4.3cm,angle=0}
    \caption[]{Squared wave function of the ground state in the $0^{++}$
    channel, measured on a $30^3$ lattice in the Higgs phase;
    left: $W_{x,1,2}(l)+W_{x,2,1}(l)$,
    right: $S_{x,1}(l)+S_{x,2}(l)$; 
    $l=0,\ldots, 15$}
    \label{fig:30b126a_10+}
    \end{center} 
  \end{minipage}
\end{figure}

\begin{figure}[!htb] 
  \begin{minipage}{8.8cm}
    \begin{center}
      \epsfig{file=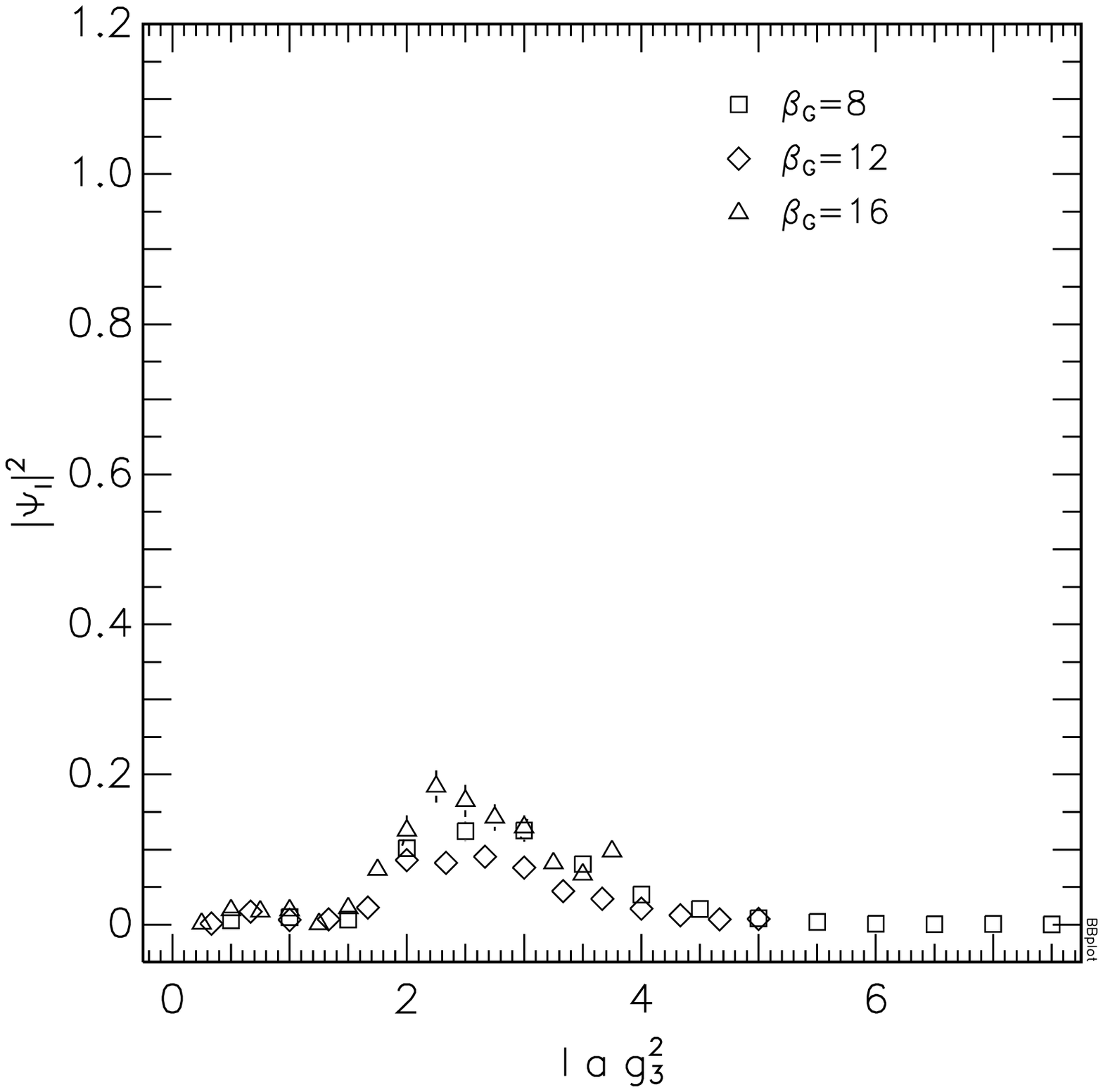,width=4.3cm,height=4.3cm,angle=0}
      \epsfig{file=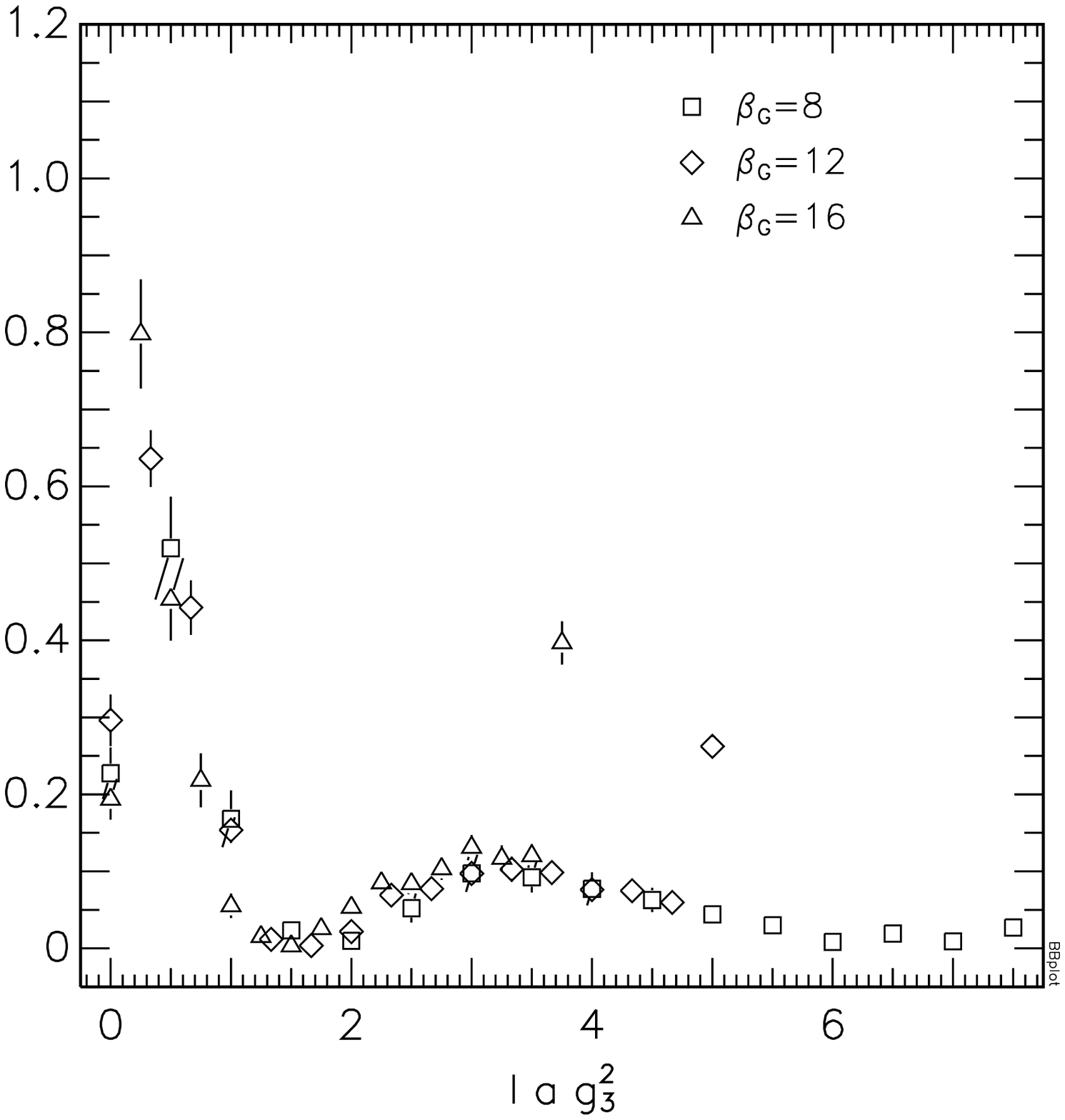,width=4.1cm,height=4.3cm,angle=0}
    \caption[]{Same as Fig.~\ref{fig:30b126a_10+} for the first excited state}
    \label{fig:30b126a_20+} 
    \end{center} 
  \end{minipage}
\end{figure}

In the $0^{++}$ channel, however, our operator set is sufficient to observe a
marked difference between the phases which is not in accordance to naive
expectations.  As a characteristic feature we observe the mixing between the
two operator types, $W$-ball operators (pure gauge d.o.f.) and operators
projecting onto Higgs states.  The squared ground state wave function looks
similar to that in the symmetric phase, contributions from Wilson loops to the
operator projecting onto the lowest mass states are absent (Fig.
\ref{fig:30b126a_10+}).
However, already the first excited Higgs state contains a noticeable
contribution from Wilson loop operators of almost 20 percent (Fig.
\ref{fig:30b126a_20+}).
Earlier measurements at a Higgs mass of $M_H^*=35$ GeV \cite{philipsen1}
(where the phase transition is very strong) did not indicate such a mixing.
We interpret this mixing of Higgs and gauge d.o.f. as a signal of the near
endpoint of the phase transition. Deeper in the Higgs phase the contribution
from gauge degrees of freedom is expected to disappear also at this high Higgs
mass.  This tendency has been checked in our simulations at $M_H^*=100$ GeV to
be discussed in the next section.  This suggests a scenario according to which
the first and second excited states present on the symmetric side (Higgs and
$W$-ball, respectively) merge into one common state on the Higgs side. The
remaining gauge degrees of freedom are fading away from that state with
further lowering of temperature (increasing $\beta_H$).
 
\begin{figure}[!htb] 
  \begin{minipage}{8.8cm}
    \begin{center}
      \epsfig{file=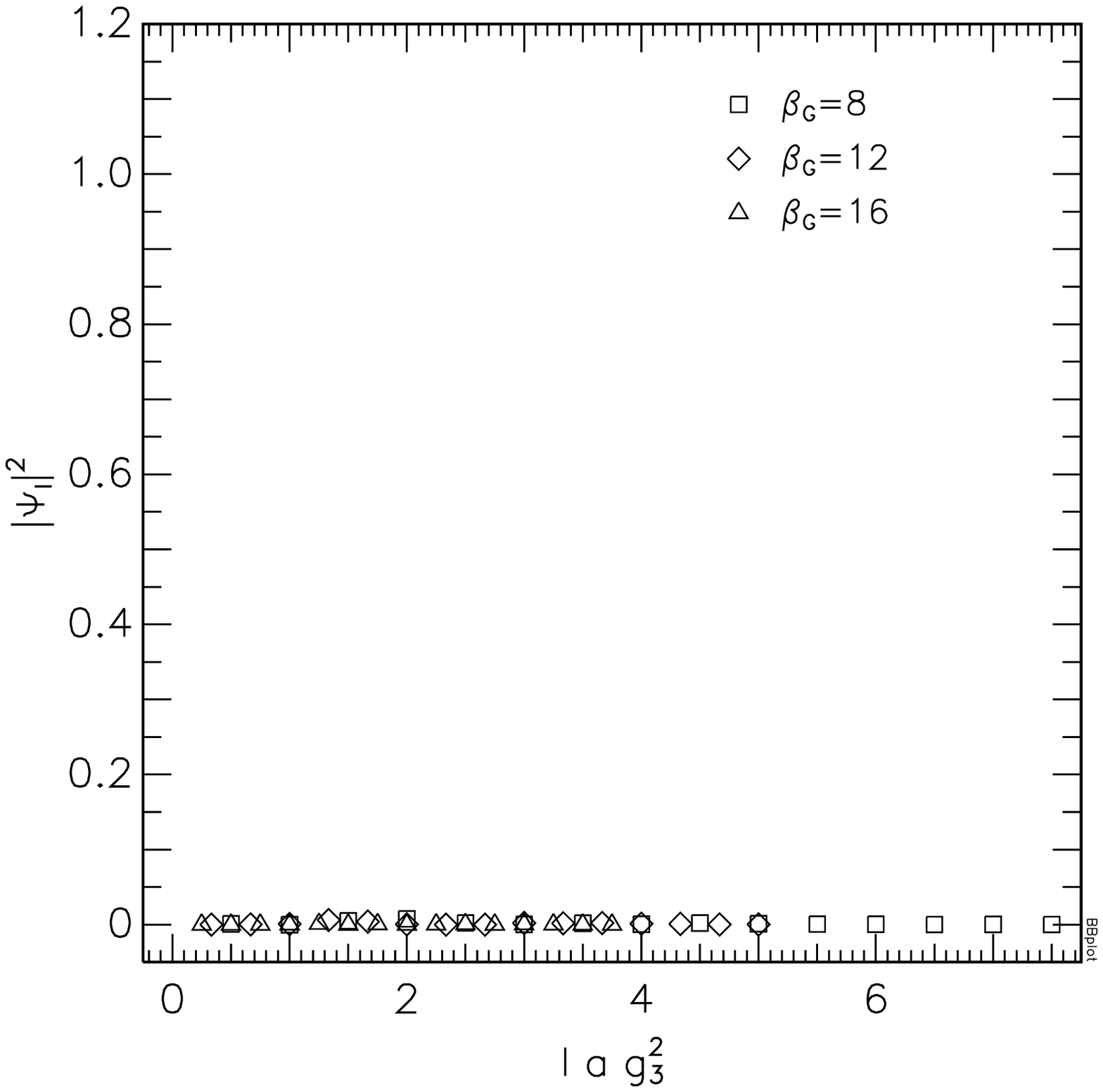,width=4.3cm,height=4.3cm,angle=0}
      \epsfig{file=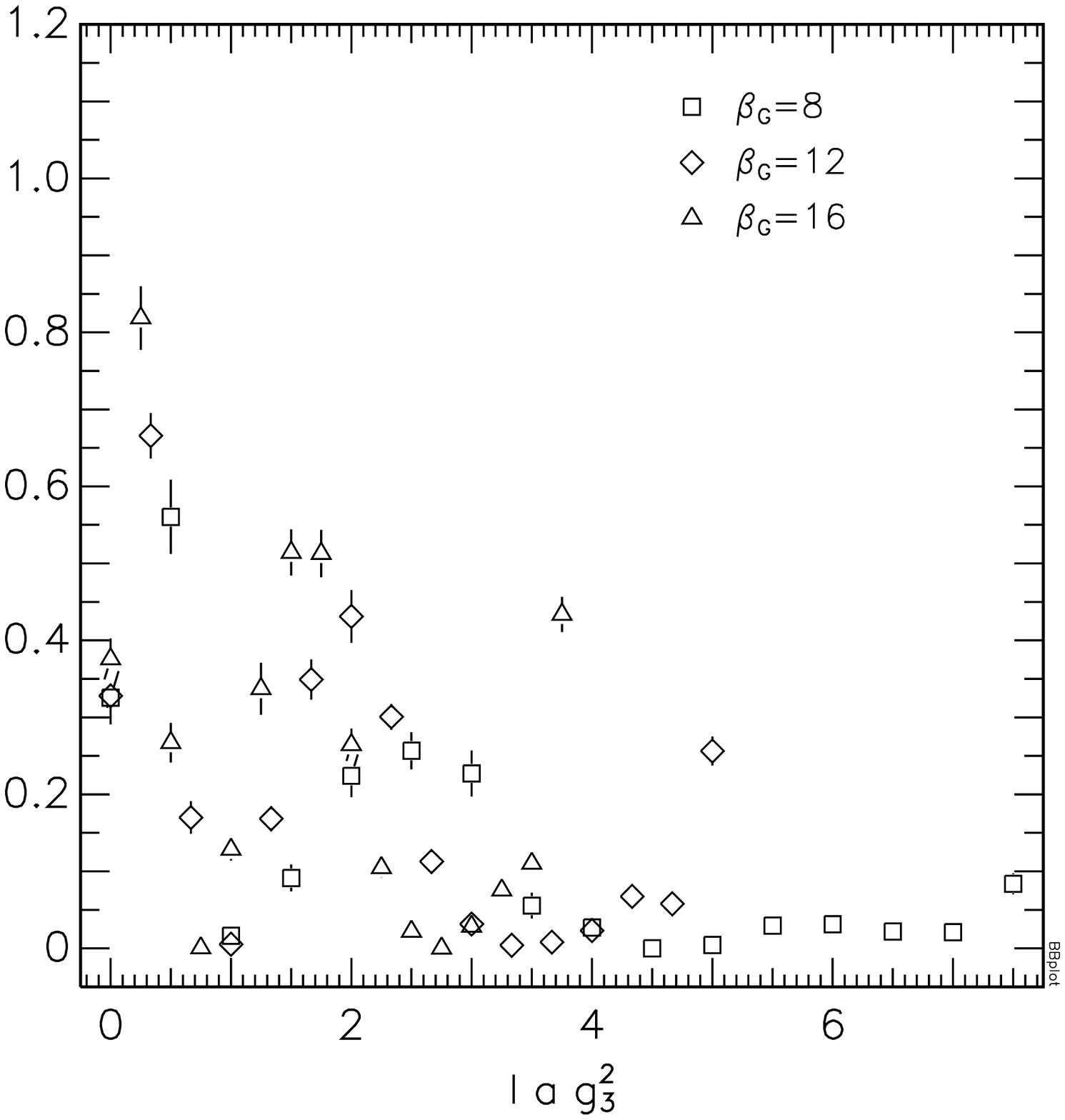,width=4.1cm,height=4.3cm,angle=0}
    \caption[]{Same as Fig.~\ref{fig:30b126a_10+} for the second excited state}
    \label{fig:30b126a_30+} 
    \end{center} 
  \end{minipage}
\end{figure}

The second excited state in the $0^{++}$ channel on the Higgs side
(Fig.~\ref{fig:30b126a_30+})
is qualitatively a state with Higgs excitations (no $W$-ball contributions)
followed by a third excited state which looks dominantly like a
$W$-ball.\footnote{Its wave function is not shown here.}

\begin{figure*}[!thb] 
  \begin{minipage}{16cm}
  \centering 
    \epsfig{file=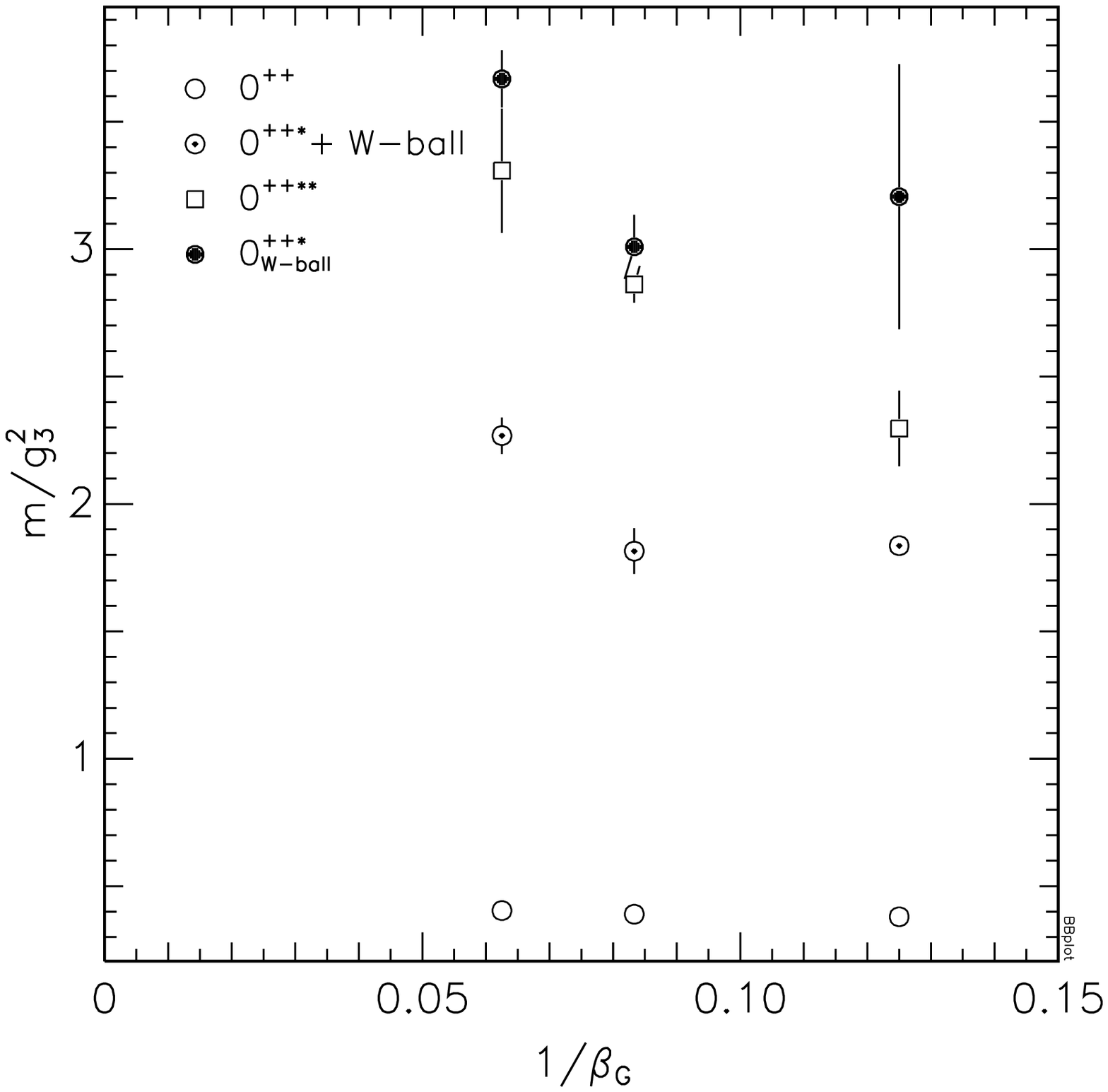,width=5cm,height=5cm,angle=0} 
    \epsfig{file=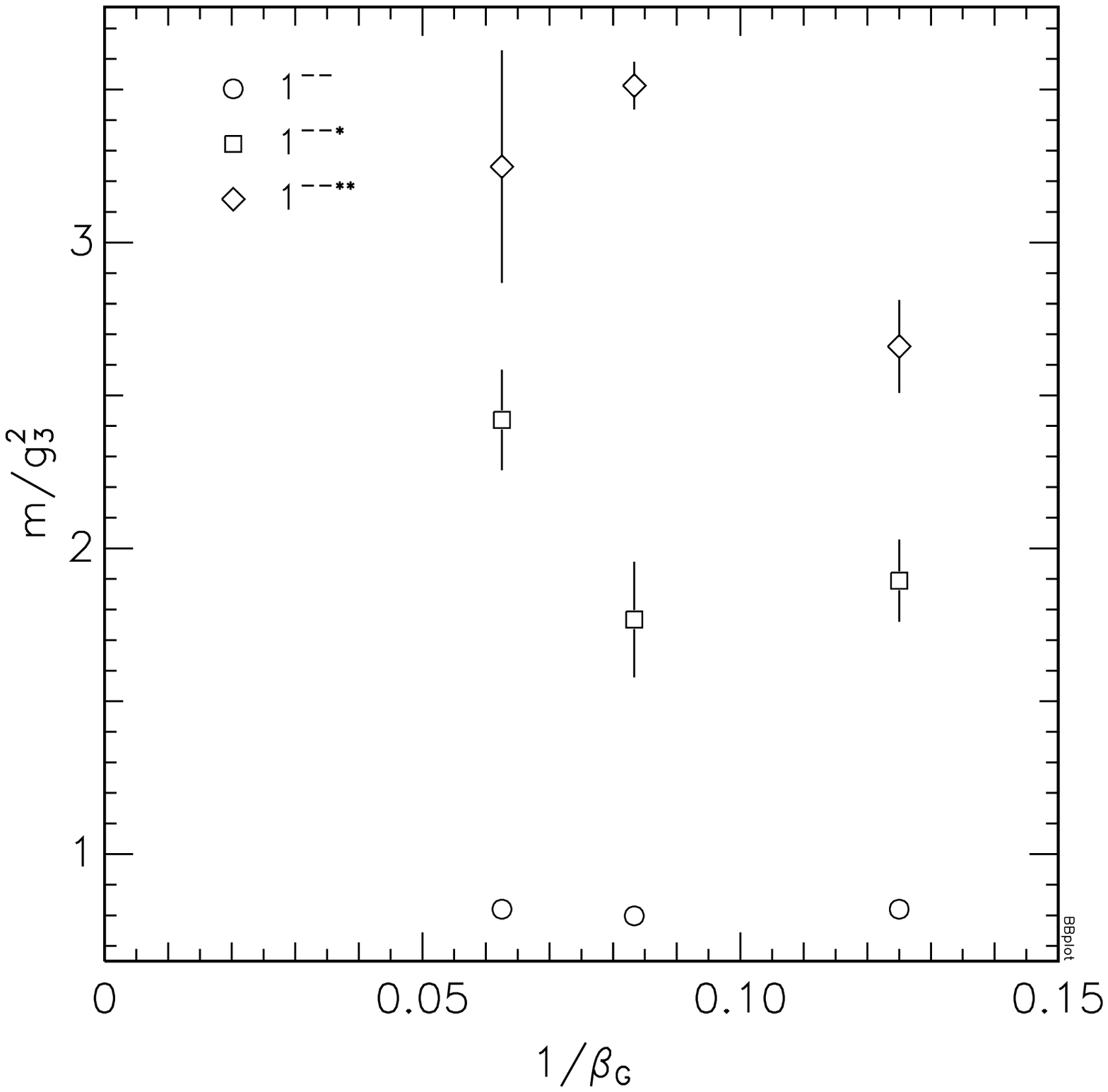,width=5cm,height=5cm,angle=0}
    \epsfig{file=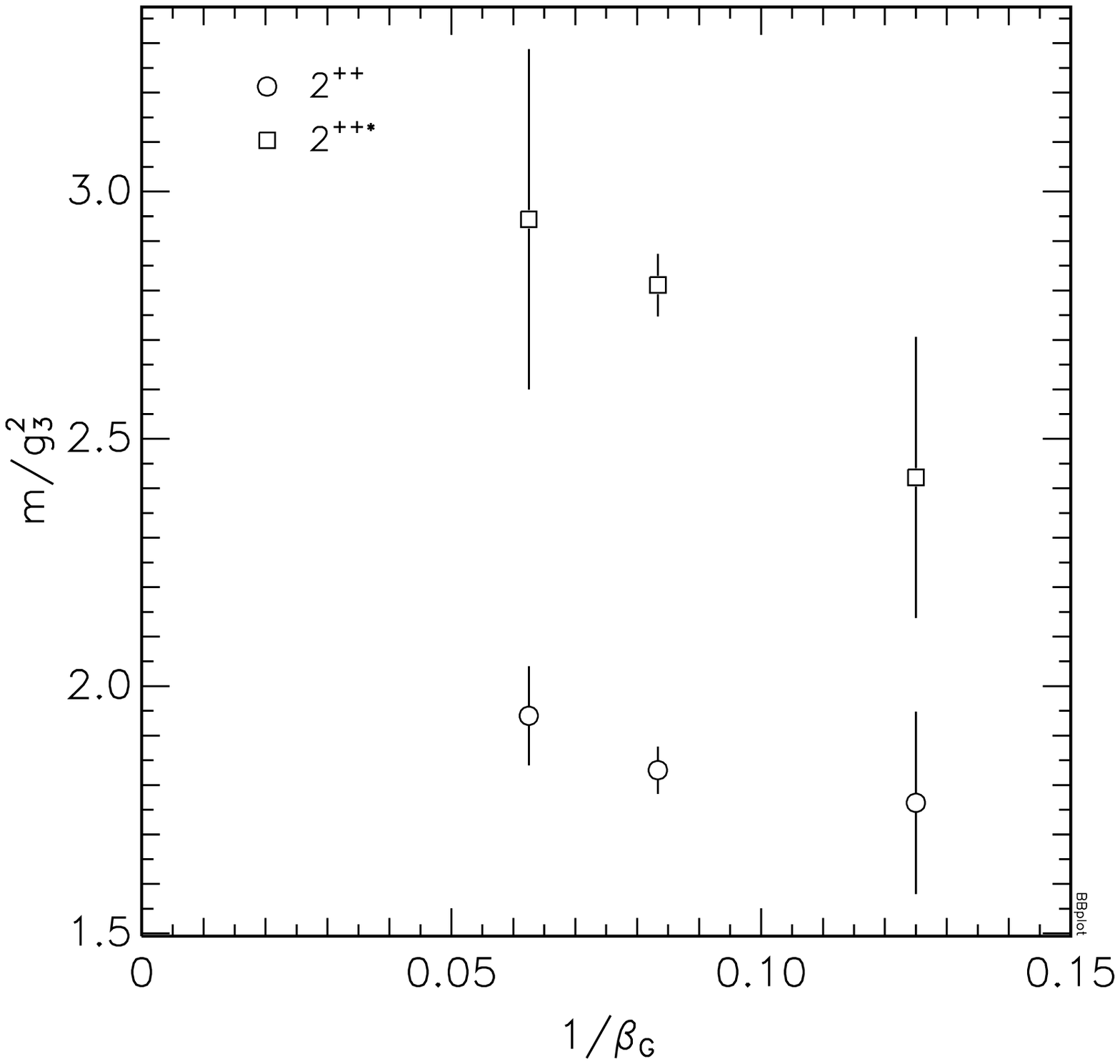,width=5cm,height=5cm,angle=0}
    \caption[]{Mass spectrum in the Higgs phase as function of 
     $1/\beta_G$}
    \label{fig:kont_spectrumh}
  \end{minipage}
\end{figure*}

In Fig.~\ref{fig:kont_spectrumh} our results for the masses at three different
gauge couplings on a line of constant physics are collected, the corresponding
values can be found in Tables \ref{tab:masskont0++h} and
\ref{tab:masskont1--2++h}.
Similar as before, the lowest mass state has a reasonable continuum limit
whereas with increasing mass the accuracy rapidly deteriorates and an
extrapolation to vanishing lattice spacing becomes difficult.
 

\section{The Spectral Change at the Rapid Crossover for 
{\boldmath $M_H^*=100$} GeV}

We have noticed that the spectra of the two phases differ mainly with respect
to the contributions of the Wilson loops (gauge d.o.f.)  to the excited states
in different channels.  Therefore we study this change in more detail while
{\it continuously} passing the crossover line (changing the hopping parameter
$\beta_H$) at fixed gauge coupling $\beta_G=12$.  Sufficiently above the phase
transition endpoint (for large enough scalar self-couplings) no large
autocorrelation times are expected which would prevent us to cross that region
of still rapidly changing thermodynamical observables and to determine the
physical excitations.  As long as a true phase transition exists, due to
tunnelling one would measure in the $0^{++}$ channel not the actual lowest
mass in the spectrum related to the respective phases but rather a
characteristic correlation length characterising the transition itself.

Our simulation have been performed at a scalar--gauge coupling ratio
$\lambda_3/g_3^2\approx 0.1953$ corresponding to the approximate Higgs mass
parameter $M_H^*=100$ GeV. We have used the same operator set as in the previous
analysis. The statistics of 4000 independent configurations per $\beta_H$
value limits the capability to analyse higher excitations. Nevertheless, we
found a behaviour very similar to our results obtained at $M_H^*=70$ GeV. 

\begin{figure*}[!thb]
   \begin{minipage}{16cm}
  \centering 
    \epsfig{file=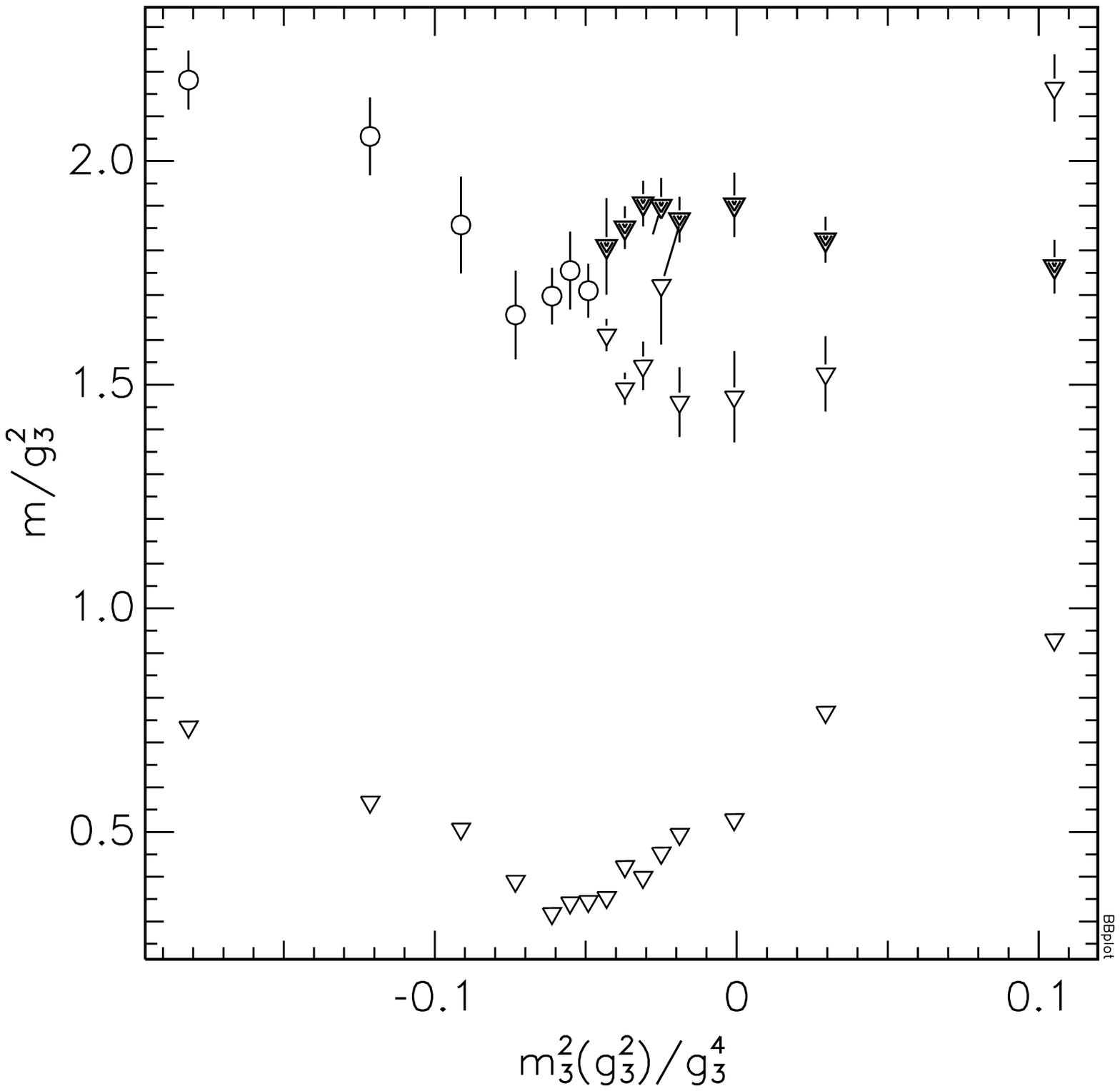,width=5cm,height=5cm,angle=0} 
    \epsfig{file=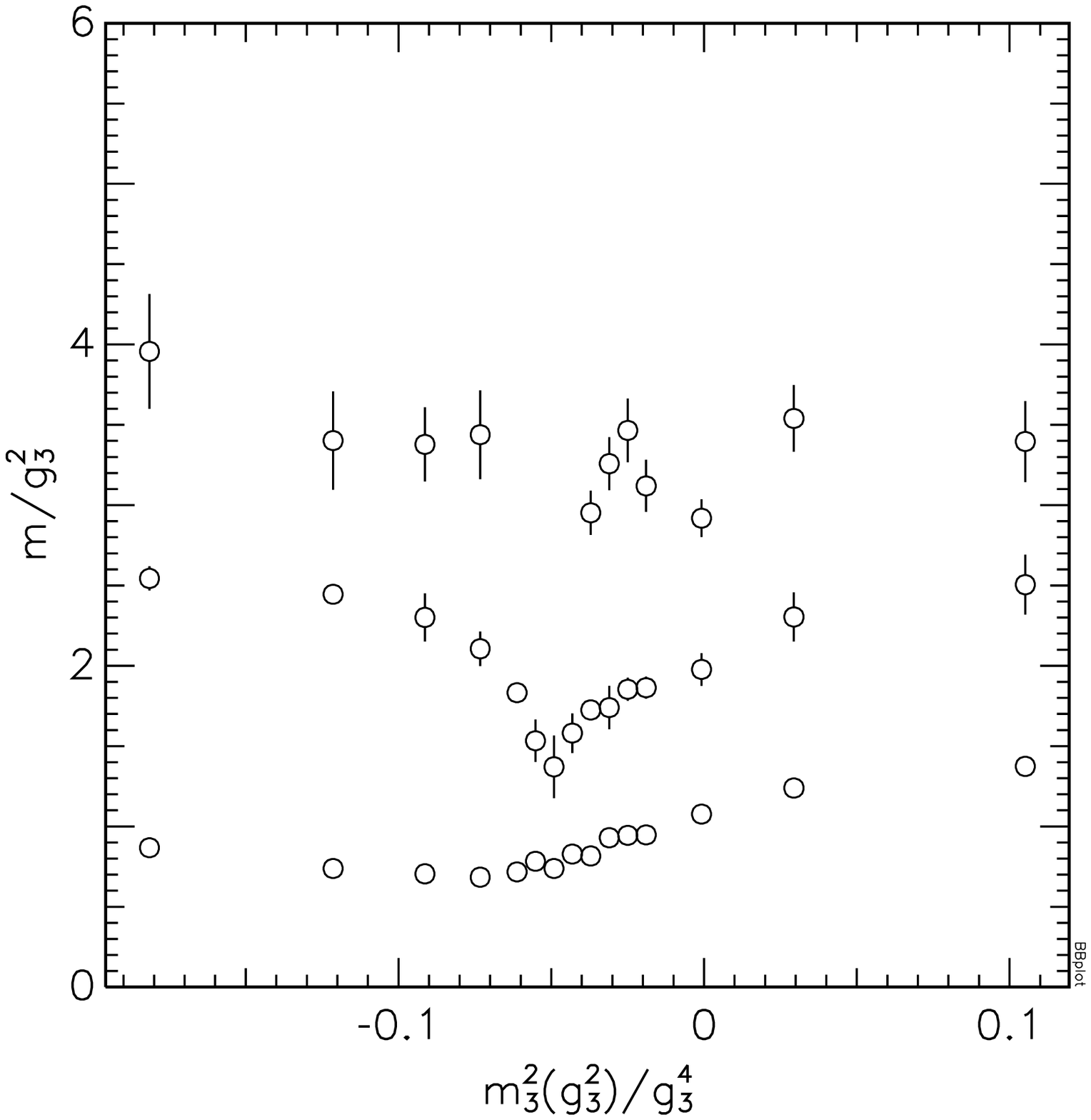,width=5cm,height=5cm,angle=0}
    \epsfig{file=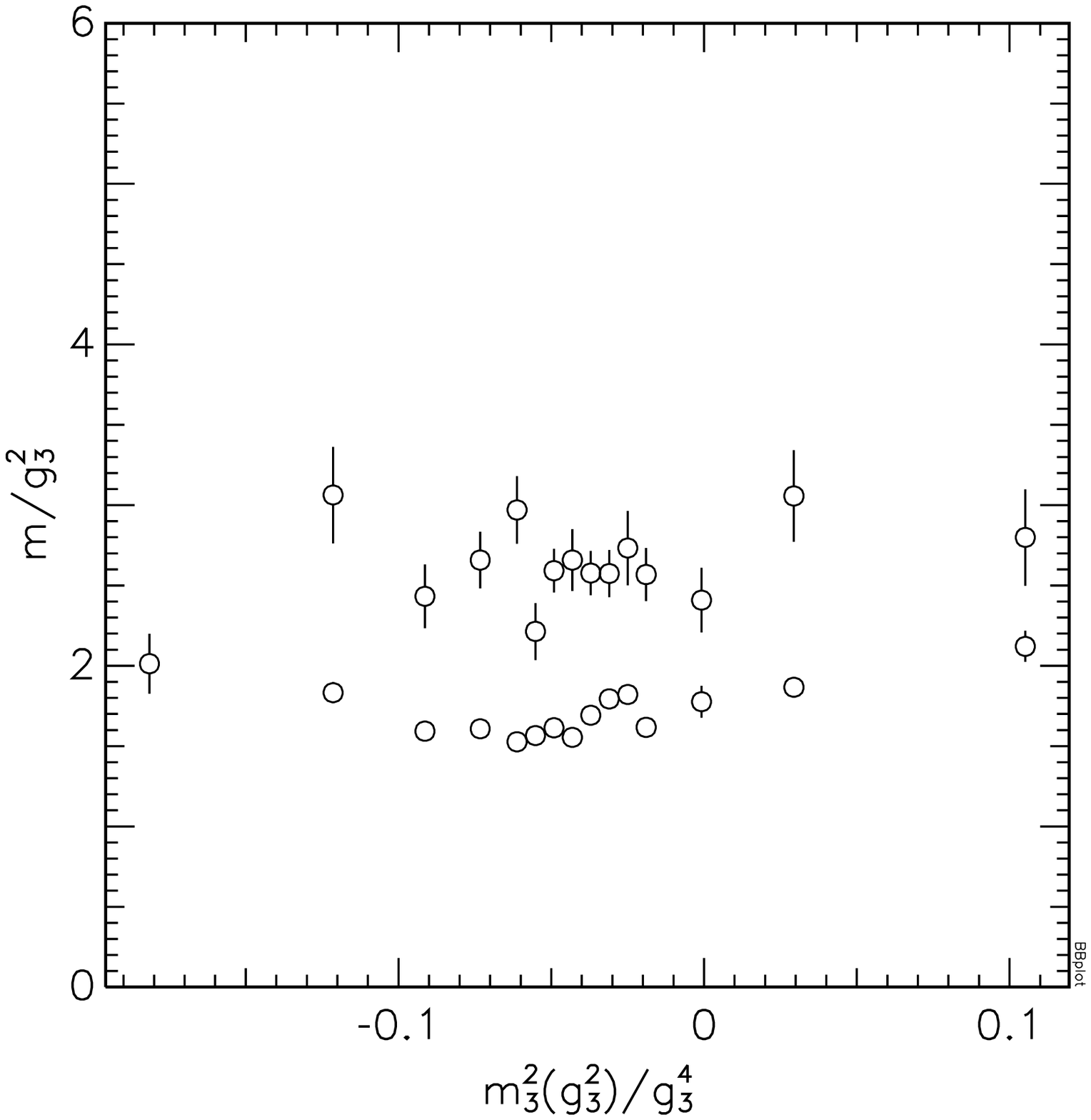,width=5cm,height=5cm,angle=0}
    \caption[]{Spectra near the crossover in the channels $0^{++}$,
     $1^{--}$ and $2^{++}$; in the $0^{++}$ channel open triangles denote 
     Higgs states, full
     triangles $W$-ball states and open circles Higgs states 
     with an admixture of
     excited gauge d.o.f.}
    \label{fig:mphi_0++1--2++}
  \end{minipage}
\end{figure*}

The similarities concern both the high temperature side of the crossover
(where one expects thermodynamic properties being close to those of the
symmetric phase at smaller Higgs mass) and the region very near to the
crossover line on the so-called Higgs side of the crossover (resembling the
Higgs phase at slightly lower Higgs mass).  In Figs.~\ref{fig:mphi_0++1--2++}
we present the spectrum of the lowest states in the $0^{++}$, $1^{--}$ and
$2^{++}$ channels as function of $m_3^2(g_3^2)/g_3^4$ (or $\beta_H$) over a
wide interval above and below the crossover ({\it i. e.} in temperature).
  
By inspecting these Figures we localize the crossover at $m_3^2(g_3^2)/g_3^4
\approx -0.05$.  Of particular interest for us is to study the mixing of Higgs
and gauge degrees of freedom in the $0^{++}$ channel.  Looking at the excited
states in this channel of Figs.~\ref{fig:mphi_0++1--2++} we conclude that the
scalar and gauge sector are approximately decoupled as long as one keeps away
from the crossover line on the high temperature side.  This has already
emphasised in Ref. \cite{philipsen,philipsen1} where cases of smaller
(characterized by the existence of a strong first order phase transition) and
larger Higgs self-couplings (far above the endpoint of the phase transition)
have been studied.  The mass of the lowest $W$-ball state (full triangle) is
roughly independent of $\beta_H$ as long as one does not come too close to the
crossover.  Thus, on the high temperature side of the crossover, the ordering
of states qualitatively resembles the spectrum at smaller values of Higgs
self-coupling (at $M_H^*=70$). If one approaches the crossover temperature the
mass of the first (Higgs-like) excitation is moving up towards the lowest
$W$-ball state whose mass decreases. At some point we observe a growing
admixture to the Higgs excitation by contributions from Wilson loop operators
which reaches about $17$ \%.  Then the wave function of the remaining state
looks similar to figure \ref{fig:30b126a_20+}.  At higher $\beta_H$ (lower
temperature, deeper in the would-be Higgs phase) the admixture from pure gauge
d.o.f. disappears again from this state.  This has been explicitly checked for
the first excited state in the $0^{++}$ channel corresponding to the data
point at $\beta_H=0.37, m_3^2(g_3^2)/g_3^4\approx -3.4$ in table \ref{t:0++}
(far on the Higgs-side) where we did not observe any contribution from Wilson
loops.

 
\section{Summary and Conclusions}

In this paper we complemented our numerical study of the electroweak phase
transition using the $3D$ $SU(2)$--Higgs model by an investigation of the
ground state and some excited states in three $J^{PC}$ channels.  Our interest
was focused on Higgs screening masses slightly below and above the end of the
first order phase transition where we wanted to understand the qualitative
relations between the higher and lower temperature regions.  This
investigation was done with the help of cross correlation matrix functions
between different operators with the same quantum numbers.  We were able
significantly to improve the old ground state mass estimates which were
obtained just by fitting a single correlation function over the plateau region
of the effective mass.  In our earlier work, the proper extension of the
operators had been selected by hand, according to a signal {\it vs.} noise
criterion for various {\it diagonal} correlators.  Implicitly this procedure,
too, gives some information concerning the best-projecting operator that
relates the vacuum state to the lowest mass bound state in the given channel.
Today, many techniques are known allowing to construct source operators with
improved projection properties.  For the $SU(2)$--Higgs and the pure $SU(2)$
model these are smearing techniques which were investigated and used in our
present context in Refs.  \cite{teper,philipsen,philipsen1}. These improved
operators take the extended character of the states into account, too,
although in a summary way.

Using the cross correlation technique the extended character can be specified
in a more detailed and systematic way.  Within a preselected set of operators,
the method finds the projection strength of a each operator onto the first few
states with quantum numbers compatible to the operators admitted. The result
can be represented as a wavefunction. Moreover, different Fock space
components (Higgs bound states and gauge-ball states) can be treated in
parallel. In the present paper this feature has been proven essential for the
understanding of excited states and to point out similarities and differences
between neighbouring regions of the extended phase diagram. Our operators were
classified in a very simple way (extended Higgs strings and Wilson loops
without internal structure) with a clear reference to the corresponding length
in lattice units.  This would allow us to interpret the results in terms of a
configuration space wave function.  Anticipating this possibility and just to
make the first step, our operator choice was the simplest one to incorporate
the notion of spatial extension.  It was possible to see the particular
effects that generically occur when the admitted operator set becomes too
small (in attempts to go too far towards the continuum limit $a\to 0$) to
cover the state under discussion.  The continuum limit of the screening masses
and wave functions has to be accompanied by correspondingly larger lattices
with the same physical volume. But then, simply increasing the operator set
simultaneously, would require much higher statistics to reach the necessary
accuracy and stability in the diagonalisation procedure. This is a reason to
include, in a next step of improvement of the method, smearing of the fields
entering the operators and/or the construction of blocked operators.

Our first measurements presented here were done at a Higgs mass of $M_H^*=70$
GeV, near to but still below the endpoint of the transition. We found that the
physics at this point is already influenced by the endpoint. The $W$-ball
states characteristic for the symmetric phase do not disappear in the Higgs
phase as long as one keeps near to the transition line.  However, the first
$W$-ball state appears here as an admixture to the first excited Higgs state.
A second $W$-ball state could also be seen in the Higgs phase very near to the
phase transition.

Our measurements in the crossover region have shown that also there the
$W$-ball state will finally disappear deeper on the low temperature side of
the crossover.  In our simulations at $M_H^*=100$GeV we could study the mixing
of the first $W$-ball with the first Higgs excited state, both being well
separated on the symmetric side of the crossover (similar to the symmetric
phase at lower Higgs mass), in some detail by continuously monitoring the
spectral evolution across the crossover line.  Over most of the parameter
space on the high temperature side of the crossover region, however,
decoupling of the gauge and scalar sector has been established.

\begin{appendix}
\renewcommand{\dblfloatpagefraction}{.1}
\renewcommand{\dbltopfraction}{0.5}

\section{Angular Momentum in 2+1 Dimensions}
\label{quantumnumbers}

Gauge invariant operators can be constructed in the $3D$ $SU(2)$--Higgs model
and can be used to define states and wave functions.  They are classified with
respect to the quantum numbers angular momentum $J$, parity $P$ and charge
conjugation $C$.  While the classification according to $C$ and $P$ is trivial
(the eigenvalues of $C$ and $P$ have the same sign for operators which live in
the same $x-y$-plane in $2+1$ dimensions) the classification according to $J$
requires some care due to the discreteness of the rotation group.  The results
are known, we collect here the main results for the convenience of the reader
and to apply them to the operators of interest.  For the $3+1$ dimensional
case the classification can be found in Refs. \cite{quantenzahlen,evertz}.
Some of the results are already described in \cite{hasenbusch}, we follow the
description in \cite{hamermesh}.

To define the angular momentum of a given operator on the lattice the
transformation properties with respect to the discrete lattice symmetries have
to be studied.  In our case the continuous symmetry group is $SO(2)$ which
gets restored in the continuum limit. In that case the angular momentum is
represented by the irreducible representations (IR) of $SO(2)$.  On the cubic
lattice the group of two-dimensional rotations and reflections is the dihedral
group ${\mathcal{D}}_4$.  This group is non-Abelian and has 8 elements ($d=8$)
and 5 irreducible representations.  The conjugation classes $C$ with their
number of elements $n_C$ and the symmetry transformations are collected in
Table \ref{t:dieder}.
\begin{table}[!htb]
  \caption[]{Symmetry transformations of   ${\mathcal{D}}_4$ group}
  \label{t:dieder}
  \begin{center}
    \begin{tabular}{|lll|} \hline
      &&\\[-2ex]
      $C(n_C)$ & & angle\\[0.5 ex] \hline
      &&\\[-2ex]
      {\boldmath$1$}(1) &  Identity  &\\[0.5 ex]
      $C_4^2(1)$        &  rotation (0,0)   &$\phi=\pi$\\[0.5 ex]
      $C_4(2)$          &  rotation (0,0)   &$\phi=\pi/2$\\[0.5 ex]
      $C_2(2)$          &  reflection x, y  &\\[0.5 ex]
      $C'_2(2)$         &  reflection diagonal    &       \\[0.5 ex] \hline
    \end{tabular}
  \end{center}
\end{table}
The corresponding IR's of ${\mathcal{D}}_4$ are denoted by $A_1$, $A_2$,
$B_1$, $B_2$ and $E$, the first four are one-dimensional, the last one has
dimension 2.
 
The characters of ${\mathcal{D}}_4$ are computed by reducing the characters of
the full $SO(2)$ group (including reflections) (table~\ref{tab:chi03}) to
those of the subduced representation $D^J_0$ by taking the discrete angles of
table \ref{t:dieder}. The result is shown in table~\ref{tab:chi23}.
\begin{table}[!htb]
    \caption[]{Characters $\chi^J$ of   full $SO(2)$ symmetry 
      $(C(\phi))$ including reflections ($\sigma_v$) for various angular
      momenta $J$}
    \label{tab:chi03}
  \begin{center}
    \begin{tabular}{|l||r|r|r|r|r|r|} \hline
      &&&&&\\[-2 ex]
      $ C_\infty\backslash J$  & 0 & 1 & 2 & 3 & 4   \\[0.5 ex] \hline \hline 
      &&&&&\\[-2 ex]
      {\boldmath$1$}          & 1 & 2         & 2          & 2 & 2  \\[0.5 ex] \hline 
      &&&&&\\[-2 ex]
      $C(\phi)$   & 1 & 2cos$\phi$  & 2cos2$\phi$ &2cos3$\phi$
      &2cos4$\phi$ \\[0.5 ex] \hline
      &&&&&\\[-2 ex]
      $\sigma_v$  & 1 & 0         & 0     & 0& 0    \\[0.5 ex] \hline
    \end{tabular}
  \end{center}
\end{table}
\begin{table}[!htb]
    \caption[]{Characters $\chi_C^J$ of  conjugation classes
      $C$ in the subduced representation $D^J_0$ }
    \label{tab:chi23}
  \begin{center}
    \begin{tabular}{|l||r|r|r|r|r|r|} \hline
      &&&&&\\[-2 ex]
      $C \backslash J$  & 0 & 1 & 2 & 3 & 4\\[0.5 ex] \hline \hline 
      &&&&&\\[-2 ex]
      {\boldmath$1$}(1)         & 1 & 2 & 2 & 2 & 2 \\[0.5 ex] \hline 
      &&&&&\\[-2 ex]
      $C_4^2(1)$     & 1 & -2& 2 & -2& 2 \\[0.5 ex] \hline
      &&&&&\\[-2 ex]
      $C_4(2)$       & 1 & 0 &-2 & 0 & 2 \\[0.5 ex] \hline
      &&&&&\\[-2 ex]
      $C_2(2)$       & 1 & 0 & 0 & 0 & 0 \\[0.5 ex] \hline
      &&&&&\\[-2 ex]
      $C'_2(2)$      & 1 & 0 & 0 & 0 & 0 \\[0.5 ex] \hline
    \end{tabular}
  \end{center}
\end{table}
The characters of the conjugation classes of ${\mathcal{D}}_4$ in the IR of
this group are taken from the character table of \cite{hamermesh} and are
shown in table~\ref{tab:chi13}.
\begin{table}[!htb]
    \caption[]{Characters $\chi_C^{IR}$ of the conjugation class $C$ in
      the
      IR of ${\mathcal{D}}_4$}
    \label{tab:chi13}
  \begin{center}
    \begin{tabular}{|l||r|r|r|r|r|} \hline
      &&&&&\\[-2 ex]
      $C \backslash IR$ & $A_1$ & $A_2$ & $B_1$ & $B_2$ & $E$\\[0.5 ex] \hline
      \hline
      &&&&&\\[-2 ex]
      {\boldmath$1$}(1)& 1  & 1  & 1  & 1  & 2  \\[0.5 ex] \hline
      &&&&&\\[-2 ex]
      $C_4^2(1)$    & 1  & 1  & 1  & 1  & -2 \\[0.5 ex] \hline
      &&&&&\\[-2 ex]
      $C_4(2)$      & 1  & 1  & -1 & -1 & 0  \\[0.5 ex] \hline
      &&&&&\\[-2 ex]
      $C_2(2)$      & 1  & -1 & 1  & -1 & 0 \\[0.5 ex] \hline
      &&&&&\\[-2 ex]
      $C'_2(2)$     & 1  & -1 & -1 &  1 & 0  \\[0.5 ex] \hline
    \end{tabular}
  \end{center}
\end{table}
\begin{table}[!htb]
    \caption[]{Multiplicities $m_J^{IR}$ of IR of ${\mathcal{D}}_4$
      in the  subduced representation  $D^J_0$ for angular momentum $J$}
    \label{tab:chi33}
  \begin{center}
    \begin{tabular}{|l||r|r|r|r|r|} \hline
      &&&&&\\[-2 ex]
      $J\backslash IR $  & $A_1$ & $A_2$ & $B_1$ & $B_2$ & $E$ \\[0.5 ex]
      \hline \hline
      &&&&&\\[-2 ex]
      0  & 1  & 0  & 0  & 0  & 0 \\[0.5 ex] \hline
      &&&&&\\[-2 ex]
      1  & 0  & 0  & 0  & 0  & 1 \\[0.5 ex] \hline
      &&&&&\\[-2 ex]
      2  & 0  & 0  & 1  & 1  & 0 \\[0.5 ex] \hline
      &&&&&\\[-2 ex]
      3  & 0  & 0  & 0  & 0  & 1 \\[0.5 ex] \hline
      &&&&&\\[-2 ex]
      4  & 1  & 1  & 0  & 0  & 0 \\[0.5 ex] \hline
    \end{tabular}
  \end{center}
\end{table}
Using the characters given in tables~\ref{tab:chi23} and \ref{tab:chi13} we
can compute the multiplicities for the IR of the dihedral group
${\mathcal{D}}_4$ for angular momentum $J$:
\begin{eqnarray}
  m_J^{IR}=\frac{1}{d}\sum\limits_C n_C \chi^{IR}_C \chi^J_C \,,
  \label{mult}
\end{eqnarray}
they are shown in table~\ref{tab:chi33}.

The multiplicities for a given operator are obtained by inspecting its
properties under those discrete transformations. The extended operators (given
in (\ref{extended})) transform as vectors with positive parity
($S_{x,\mu}(l)$) or with negative parity ($V_{x,\mu}^b(l)$). The characters
and the multiplicities are shown in table \ref{tab:ca3}.
\begin{table}[!thb]
    \caption[]{Characters and  multiplicities for the operators
      $S_{x,\mu}(l)$ and $V_{x,\mu}^b(l)$}
    \label{tab:ca3}
  \centering
  \begin{tabular}{|l|c|c|c|c|c|} \hline
      &&&&&\\[-2 ex]
    $\chi_C^{IR}$& {\boldmath$1$}(1)& $C_4^2(1)$    &  $C_4(2)$ &  $C_2(2)$ &
    $C'_2(2)$  \\[0.5 ex] \hline
      &&&&&\\[-2 ex]
    $S_{x,\mu}(l)$  & 2   & 2          &  0 &  2        & 0 \\[0.5 ex] \hline
      &&&&&\\[-2 ex]
    $V_{x,\mu}^b(l)$ &  2   & -2       &  0 & 0        & 0 \\[0.5 ex] \hline
 \hline
      &&&&&\\[-2 ex]
    $m_J^R$  & $A_1$ & $A_2$ & $B_1$ & $B_2$ & $E$  \\[0.5 ex] \hline
      &&&&&\\[-2 ex]
    $S_{x,\mu}(l)$  & 1  & 0  & 1  & 0  & 0 \\[0.5 ex] \hline
      &&&&&\\[-2 ex]
    $V_{x,\mu}^b(l)$& 0 & 0  & 0  & 0  & 1 \\[0.5 ex] \hline
  \end{tabular}
\end{table} 
We observe that the operator $S_{x,\mu}(l)$ transforms simultaneously under
the IR $A_1$ and $B_1$, whereas $V_{x,\mu}^b(l)$ transforms with $E$ only.

Using a projection operator $P^{IR}$ one is able to project onto \it one \rm
irreducible representation in the case of $S_{x,\mu}(l)$.  A projection
operator for a given IR is constructed as the sum over all conjugation classes
$C$ weighted by the character $\chi_C^{IR}$ for that IR
(table~\ref{tab:chi13}):
\begin{equation}
  P^{IR}=\sum\limits_C \chi_C^{IR} C \,.
\end{equation}
The $C$ have to be taken in a matrix representation taking into account
parity.

For the operator $S_{x,\mu}(l)$ the projection matrices can be used in the
form:
\begin{eqnarray}
  P^{A_1}=4 \left ( 
    \begin{array}{rr} 
      1 & 1 \\ 
      1 & 1 \\
    \end{array} 
  \right ) , \ \ \ 
  P^{B_1}=4 \left ( 
    \begin{array}{rrr}
      1 & -1 \\
      -1&  1 \\  
    \end{array} 
  \right ).
\end{eqnarray}
From here the fixed angular momenta of the operators to be used are defined as
shown in table~\ref{tab:operators}.
\begin{table}[!htb]
    \caption[]{Quantum numbers and gauge invariant lattice operators}
    \label{tab:operators}
  \begin{center}
    \begin{tabular}{|rl|} \hline
      &\\[-2 ex]
      $0^{++}$: & $S_{x,1}(l)+S_{x,2}(l)$ \\[0.5 ex]
      &\\[-2 ex]
      & $W_{x,1,2}(l)+W_{y,1,2}(l)$\\[0.5 ex] \hline 
      &\\[-2 ex]
      $1^{--}$:  & $V_{x,1}^b(l)+V_{x,2}^b(l)$\\[0.5 ex] \hline
      &\\[-2 ex]
      $2^{++}$:  & $S_{x,1}(l)-S_{x,2}(l)$\\[0.5 ex] 
      &\\[-2 ex]
      & $W_{x,1,2}(l)-W_{y,1,2}(l)$  \\[0.5 ex]
      \hline
    \end{tabular}
  \end{center}
\end{table}
It is obvious from table \ref{tab:chi33} that all operators describing a $J=0$
angular momentum state also overlap to a $J=4$ state. These higher states are
expected to be sufficiently heavier than the lightest $J=0$ state.  Similarly
this also holds for the other operators.  Thus for the cubic lattice only
operators with $J ({\mathrm {mod}} \;4)$ can be constructed what restricts the
identification of highly excited states in the channels discussed here.

\onecolumn
  \section{Tables of Masses for Ground States and
    Excitations}
  \label{appB}
  \begin{table*}[!thb]
    \begin{minipage}{10cm}
      \caption[]{Masses of $0^{++}$ states 
        in units of $g_3^2$
        at $M_H^*=70$~GeV 
        in the symmetric phase 
        ($30^3$ lattice)}
      \label{tab:masskont0+}
    \end{minipage}

    \begin{tabular}{|r|r|l|l|l|l|l|} \hline
      &&&&&&\\[-2 ex]
      $\beta_G$ & $\beta_H$ & $0^{++}$& $0^{++*}$&
      $0^{++}_{\mathrm{W-ball}}$ &  $0^{++**}$& 
      $0^{++*}_{\mathrm{W-ball}}$ \\[0.5 ex] \hline
      &&&&&&\\[-2 ex]
      8  & 0.3490 & 0.504(16)& 1.628(64) & 1.998(37) 
      & &                \\[0.5 ex] \hline
      &&&&&&\\[-2 ex]
      12 & 0.3434 & 0.398(18)& 1.515(33) & 1.902(26) 
      & 2.679(75)& 2.868(123)\\[0.5 ex] \hline
      &&&&&&\\[-2 ex]
      16 & 0.3407 & 0.404(59)& 1.824(100)& 2.080(83)
      & 3.604(172) & 2.736(228)\\[0.5 ex] \hline 
    \end{tabular}
  \end{table*}
  \vspace{-1cm}
  \begin{table*}[!thb]
    \begin{minipage}{10cm}
      \caption[]{Masses of $1^{--}$ and $2^{++}$ states  
        in units of $g_3^2$
        at $M_H^*=70$~GeV 
        in the symmetric phase 
        ($30^3$ lattice)}
      \label{tab:masskont1--2++}
    \end{minipage}

    \begin{tabular}{|r|r|l|l|l|l|} \hline
      &&&&\\[-2 ex]
      $\beta_G$  &    $\beta_H$     & $1^{--}$  & $1^{--*}$  
      & $2^{++}$\\[0.5 ex] \hline
      &&&&\\[-2 ex]
      8  & 0.3490 & 1.122(35)& 2.054(87) & 1.498(68)\\[0.5 ex] \hline 
      &&&&\\[-2 ex]
      12 & 0.3434 & 1.074(14)& 1.857(68) & 1.518(78)\\[0.5 ex] \hline 
      &&&&\\[-2 ex]
      16 & 0.3407 & 0.992(68)& 2.028(259)& 1.788(21)  \\[0.5 ex] \hline
    \end{tabular}
  \end{table*}
  
  \vspace{-1cm}
  \begin{table*}[!thb]
    \begin{minipage}{10cm}
    \caption[]{Masses of $0^{++}$ states 
      in units of $g_3^2$
      at $M_H^*=70$~GeV 
      in the Higgs phase
      ($30^3$ lattice)}
    \label{tab:masskont0++h}
    \end{minipage}

    \begin{tabular}{|r|r|l|l|l|l|l|} \hline
      &&&&&\\[-2 ex]
      $\beta_G$ & $\beta_H$ & $0^{++}$& $0^{++*}$ {\small + gauge d.o.f.}
      &
      $0^{++}_{\mathrm{W-ball}}$ &  $0^{++**}$ \\[0.5 ex] \hline
      &&&&&\\[-2 ex]
      8  & 0.34970 & 0.380(4)& 1.836(36) & 2.296(148) & 3.206(519)\\[0.5 ex]
      \hline
      &&&&&\\[-2 ex]
      12 & 0.34368 & 0.390(3)& 1.815(90) & 2.862(71) & 3.009(126)\\[0.5 ex]
      \hline
      &&&&&\\[-2 ex]
      16 & 0.34088 & 0.404(4)& 2.268(71)& 3.308(244) & 3.668(112)\\[0.5 ex]
      \hline 
    \end{tabular}
  \end{table*}
  \vspace{-1cm}
  \begin{table*}[!thb]
    \begin{minipage}{10cm}
    \caption[]{Masses of $1^{--}$ and $2^{++}$ states  
      in units of $g_3^2$
      at $M_H^*=70$~GeV 
      in the Higgs phase
      ($30^3$ lattice)}
    \label{tab:masskont1--2++h}
    \end{minipage}

    \begin{tabular}{|r|r|l|l|l|l|l|} \hline
      &&&&&\\[-2 ex]
      $\beta_G$  &    $\beta_H$     & $1^{--}$  & $1^{--*}$ 
      & $2^{++}$ & $2^{++*}$ \\[0.5 ex] \hline
      &&&&&\\[-2 ex]
      8  & 0.34970 & 0.820(8)& 1.894(134) & 1.764(184) & 2.422(288)\\[0.5 ex]
      \hline
      &&&&&\\[-2 ex]
      12 & 0.34368 & 0.798(6)& 1.767(189) & 1.830(48)  & 2.811(63)\\[0.5 ex]
      \hline 
      &&&&&\\[-2 ex]
      16 & 0.34088 & 0.820(12)&2.420(151) & 1.940(100) & 2.944(344)\\[0.5 ex]
      \hline
    \end{tabular}
  \end{table*}
  \vspace{-1cm}
  
  \begin{table*}[!thb]
    \begin{minipage}{10cm}
    \caption[]{Masses of $0^{++}$ states 
      in units of $g_3^2$
      at $M_H^*=100$~GeV (crossover region) 
      depending on $\beta_H$} 
    \label{t:0++}
    \end{minipage}

    \begin{tabular}{|c|r|l|l|l|} \hline
      &&&&\\[-2 ex]
      & $\beta_H$ & $0^{++}$& $0^{++*}$ &$0^{++}_{\mathrm{W-ball}}$ \\[0.5 ex]
      \hline
      &&&&\\[-2 ex]
      symmetric       & 0.34200 & 2.037(51)& 3.330(225) & 1.770(59)  \\[0.5 ex]
      \cline{2-5}
      &&&&\\[-2 ex]
      & 0.34510 & 0.930(18)& 2.163(75) & 1.764(59)  \\[0.5 ex]
      \cline{2-5}
      &&&&\\[-2 ex]
      & 0.34560 & 0.768(12)& 1.525(84) & 1.824(51)  \\[0.5 ex]
      \cline{2-5}
      &&&&\\[-2 ex]
      & 0.34580 & 0.528(14)& 1.473(102)& 1.902(71)  \\[0.5 ex]
      \cline{2-5}
      &&&&\\[-2 ex]
      & 0.34592 & 0.495(14)& 1.461(78) & 1.869(51)  \\[0.5 ex]
      \cline{2-5}
      &&&&\\[-2 ex]
      & 0.34596 & 0.453(14)& 1.722(131)& 1.899(63)  \\[0.5 ex]
      \cline{2-5}
      &&&&\\[-2 ex]
      & 0.34600 & 0.399(53)& 1.542(53) & 1.905(51)  \\[0.5 ex]
      \cline{2-5}
      &&&&\\[-2 ex]
      & 0.34604 & 0.423(14)& 1.491(35) & 1.851(48)  \\[0.5 ex]
      \cline{2-5}
      &&&&\\[-2 ex]
      & 0.34608 & 0.354(12)& 1.611(35) & 1.809(107) \\[0.5 ex] 
      \cline{2-5}
      &&&\multicolumn{2}{l|}{}\\[-2 ex]
      &   & $0^{++}$   &\multicolumn{2}{l|}{$0^{++*}$ + gauge
        d.o.f.} \\[0.5 ex] 
      \cline{2-5}
      &&&\multicolumn{2}{l|}{}\\[-2 ex]
      & 0.34612     & 0.345(6) &\multicolumn{2}{l|}{1.710(60)} \\[0.5 ex] 
      \cline{2-5}
      &&&\multicolumn{2}{l|}{}\\[-2 ex]
      & 0.34616 & 0.342(12)&\multicolumn{2}{l|}{ 1.755(87)} \\[0.5 ex]
      \cline{2-5}
      & 0.34620 & 0.318(7) &\multicolumn{2}{l|}{ 1.698(63)} \\[0.5 ex]
      \cline{2-5}
      &&&\multicolumn{2}{l|}{}\\[-2 ex]
      & 0.34628 & 0.390(6) &\multicolumn{2}{l|}{ 1.656(99)} \\[0.5 ex]
      \cline{2-5}
      &&&\multicolumn{2}{l|}{}\\[-2 ex]
      & 0.34640 & 0.507(6) &\multicolumn{2}{l|}{ 1.857(108)} \\[0.5 ex]
      \cline{2-5}
      &&&\multicolumn{2}{l|}{}\\[-2 ex]
      & 0.34660 & 0.567(12)&\multicolumn{2}{l|}{ 2.055(87)} \\[0.5 ex]
      \cline{2-5}
      &&&\multicolumn{2}{l|}{}\\[-2 ex]
      & 0.34700 & 0.735(9) &\multicolumn{2}{l|}{ 2.181(66)} \\[0.5 ex]
      \cline{2-5}
      &&&\multicolumn{2}{l|}{}\\[-2 ex]
      &   & $0^{++}$   &\multicolumn{2}{l|}{$0^{++*}$ } \\[0.5 ex] 
      \cline{2-5}
      &&&\multicolumn{2}{l|}{}\\[-2 ex]
      Higgs &0.37000 & 2.723(48)&\multicolumn{2}{l|}{5.250(339) } \\[0.5 ex] 
      \hline
    \end{tabular}
  \end{table*}
  
  \vspace{-1cm}
  \begin{table*}[!thb]
    \begin{minipage}{10cm}
    \caption[]{Masses of $1^{--}$ and $2^{++}$ states
      in units of $g_3^2$
      at $M_H^*=100$~GeV (crossover region) 
      depending on $\beta_H$}
    \label{t:1--2++}
    \end{minipage}

    \begin{tabular}{|c|r|l|l|l|l|l|} \hline
      &&&&&&\\[-2ex]
      & $\beta_H$ & $1^{--}$& $1^{--*}$ &$1^{--**}$ & $2^{++}$& $2^{++*}$ 
      \\[0.5 ex] \hline
      &&&&&&\\[-2ex]
      symmetric       & 0.34200 & 3.261(177)& &  
      & 3.255(84) &  \\[0.5 ex] 
      \cline{2-7}
      &&&&&&\\[-2ex]
      & 0.34510 & 1.374(48)& 2.505(185) & 3.396(252)
      & 2.121(96)& 2.799(300) \\[0.5 ex] 
      \cline{2-7}
      &&&&&&\\[-2ex]
      & 0.34560 & 1.239(30)& 2.304(152) & 3.540(206)
      & 1.866(63)& 3.057(284) \\[0.5 ex] 
      \cline{2-7}
      &&&&&&\\[-2ex]
      & 0.34580 & 1.077(14)& 1.977(102) & 2.919(116)
      & 1.776(98)& 2.409(201) \\[0.5 ex] 
      \cline{2-7}
      &&&&&&\\[-2ex]
      & 0.34592 & 0.948(21)& 1.863(68)  & 3.120(162)
      & 1.670(35)& 2.568(164) \\[0.5 ex] 
      \cline{2-7}
      &&&&&&\\[-2ex]
      & 0.34596 & 0.945(21)& 1.854(71)  & 3.465(197)
      & 1.821(53)& 2.733(231) \\[0.5 ex] 
      \cline{2-7}
      &&&&&&\\[-2ex]
      & 0.34600 & 0.930(12)& 1.740(135) & 3.258(164)
      & 1.794(51)& 2.574(147) \\[0.5 ex] 
      \cline{2-7}
      &&&&&&\\[-2ex]
      & 0.34604 & 0.816(35)& 1.725(59)  & 2.952(137)
      & 1.692(45)& 2.577(137) \\[0.5 ex] 
      \cline{2-7}
      &&&&&&\\[-2ex]
      & 0.34608 & 0.828(26)& 1.581(123) &           
      & 1.554(51)& 2.658(192) \\[0.5 ex] 
      \cline{2-7}
      &&&&&&\\[-2ex]
      & 0.34612 & 0.738(24)& 1.371(194) &           
      & 1.614(29)& 2.592(135) \\[0.5 ex] 
      \cline{2-7}
      &&&&&&\\[-2ex]
      & 0.34616 & 0.783(17)& 1.533(132) &           
      & 1.566(42)& 2.214(177) \\[0.5 ex] 
      \cline{2-7}
      &&&&&&\\[-2ex]
      & 0.34620 & 0.717(32)& 1.833(60)  &           
      & 1.527(42)& 2.970(210) \\[0.5 ex] 
      \cline{2-7}
      &&&&&&\\[-2ex]
      & 0.34628 & 0.684(21)& 2.106(107) & 3.438(275)
      & 1.608(44)& 2.658(177) \\[0.5 ex] 
      \cline{2-7}
      &&&&&&\\[-2ex]
      & 0.34640 & 0.705(26)& 2.301(150) & 3.378(231)
      & 1.593(56)& 2.433(197) \\[0.5 ex] 
      \cline{2-7}
      &&&&&&\\[-2ex]
      & 0.34660 & 0.738(24)& 2.445(63)  & 3.402(305)
      & 1.833(65)& 3.063(300) \\[0.5 ex] 
      \cline{2-7}
      &&&&&&\\[-2ex]
      & 0.34700 & 0.867(12)& 2.544(75)  & 3.957(357)
      & 2.013(185)&  \\[0.5 ex] 
      \cline{2-7} 
      &&&&&&\\[-2ex]
      Higgs           & 0.37000  & 2.400(33) &   &  &  & \\[0.5 ex] 
      \hline
    \end{tabular}
  \end{table*}
\end{appendix}
\twocolumn

\end{document}